%% file: User_trust_exp.tex
\documentclass{article}
\usepackage{hyphenat}
\PassOptionsToPackage{numbers}{natbib}
\usepackage[eandd]{neurips_2026}

\usepackage[utf8]{inputenc} 
\usepackage[T1]{fontenc}    
\usepackage{hyperref}       
\usepackage{url}            
\usepackage{booktabs}       
\usepackage{amsfonts}       
\usepackage{nicefrac}       
\usepackage{microtype}      
\usepackage{xcolor}         
\usepackage{amsmath}
\usepackage{graphicx}
\usepackage{subcaption}
\usepackage{multirow}
\usepackage{wrapfig}
\usepackage{caption}
\usepackage[most]{tcolorbox}
\usepackage{xcolor}
\usepackage{fvextra}
\usepackage[most]{tcolorbox}
\usepackage{graphicx}
\usepackage{float}
\usepackage{placeins}

\newtcolorbox{prompttemplate}[2][]{
    colback=gray!5,
    colframe=black!70,
    coltitle=white,
    colbacktitle=black!70,
    fonttitle=\bfseries,
    title=#2,
    breakable,
    enhanced,
    sharp corners=south,
    boxrule=0.5pt,
    left=6pt, right=6pt, top=6pt, bottom=6pt,
    #1
}
\title{Evaluating the False Trust Engendered by LLM Explanations}

\author{
  Vardhan Palod\thanks{Equal contribution; author order decided by coin toss.} \quad
  Upasana Biswas\footnotemark[1] \quad
  Subbarao Kambhampati \\
  School of Computing \& AI, Arizona State University \\
  \texttt{\{vpalod, ubiswas2, rao\}@asu.edu}
}

\begin{document}

\maketitle

\begin{abstract}
Large Language Models (LLMs) and Large Reasoning Models (LRMs) are increasingly used for critical tasks, yet they provide no guarantees about the correctness of their solutions. Users must decide whether to trust the model's answer, aided by reasoning traces, their summaries, or post-hoc generated explanations.
These reasoning traces, despite evidence that they are neither faithful representations of the model's computations nor necessarily semantically meaningful, are often interpreted as provenance explanations. It is unclear whether explanations or reasoning traces help users identify when the AI is incorrect, or whether they simply persuade users to trust the AI regardless.
In this paper, we take a \textit{user-centered} approach and develop an evaluation protocol to study how different explanation types affect users' ability to judge the correctness of AI-generated answers and engender false trust in the users. We conduct a between-subject user study, simulating a setting where users do not have the means to verify the solution and analyze the false trust engendered by commonly used LLM explanations — reasoning traces, their summaries and post-hoc explanations. We also test a contrastive dual explanation setting where we present arguments for and against the AI's answer.
We find that reasoning traces and post-hoc explanations are persuasive but not informative: they increase user acceptance of LLM predictions regardless of their correctness. 
In contrast, dual explanation is the only condition that genuinely improves users' ability to distinguish correct from incorrect AI outputs. 
\end{abstract}
\input{sections/intro_eval}
\input{sections/related_works}

\input{sections/methods_2}
\input{sections/results_v2}

\input{sections/discussions_v2}
\section{Conclusion}
In this paper, we introduce a novel evaluation protocol to isolate the effects of different explanation conditions on user trust by simulating a setting where users do not have the ability or the means to directly verify the LLM's prediction. We conducted a between-users study where users needed to judge whether a LLM generated answer is correct. We found that reasoning traces, their summaries and post hoc LLM generated answer explanations can hinder the detection of incorrect AI answers and engender false trust in users. In an attempt to mitigate this, we evaluated a post-hoc dual explanation generation paradigm, where users are exposed to both merits and demerits of the LLM's generated answer. We find that the dual explanation condition minimizes false trust while maintaining a high detection accuracy.
\paragraph{Acknowledgements:} This research is supported in part by grants from ONR (N00014-25-1-2301 and N00014-23-1-2409), DARPA (HR00112520016), DoD RAI (via CMU subcontract 25-00306-SUB-000), and a generous gift from Qualcomm.
\bibliographystyle{unsrtnat}
\bibliography{sections/ref, sections/bs_ref}
\input{sections/appendix}

\end{document}

%% file: sections/intro_eval.tex
\section{Introduction}
With the advent of Large Language models (LLMs), the field of AI has been revolutionized. LLMs have demonstrated impressive capabilities across a wide array of tasks, from planning and reasoning to instruction following and story writing~\citep{zhao2026}. Unlike traditional AI systems, these generative language models are broad and general purpose models, leading to their rapid adoption by lay users. 
However, this adoption comes with a caveat.
Large Language Models confidently answer any question put to them, regardless of whether the answer is correct. As these systems are increasingly used for critical decisions in healthcare, education, and legal domains, two critical questions arise: how do we develop methods to engender appropriate rather than uncritical trust in end users, and how do we evaluate the effectiveness of such methods.


Traditional Explainable AI (xAI) techniques were designed for deep and narrow, task-specific AI systems that were experts in one domain~\citep{deepblue, alphago}. Since they did not share the same internal representations as humans, these systems were reliable but opaque to users. 
The goal of developing explanation techniques was to increase user trust by explaining the system's decision-making process in terms which the users can understand~\citep{rib, been}. The xAI challenge was ``how do we get users to trust a system which doesn't speak the end-user's language ?''.

LLMs change the setting quite a bit. They communicate in natural language and readily provide a palatable explanation of their generated answers, while not providing any guarantees of correctness or faithfulness~\citep{turpin, anth, chua2025deepseek, lanham2023measuring}.
We observe this in a new crop of language models called Large Reasoning Models (LRMs) which generate a sequence of intermediate tokens (colloquially or perhaps fancifully called a “Chain of Thought” or “reasoning trace”) before producing an answer sequence~\citep{deepseek}. These reasoning traces have often been interpreted as the intermediate steps which the model took to arrive at the final answer, leading the community to treat them as \textbf{provenance explanations}~\citep{sree}. However, some researchers have cautioned against interpreting reasoning traces as faithful representations of the model's internal computations~\citep{turpin}, and have demonstrated that these traces do not need to have any semantic meaning or end-user interpretability to produce correct answers~\citep{bs, int, kambhampati2026stop}. 
The frontier model makers– OpenAI~\citep{openai}, Google~\citep{gemini}, Anthropic~\citep{anthropic}- do not show the actual reasoning traces which the models produced rather expose the users to what seems like a summarized version of the reasoning traces. 
GPT-OSS models~\citep{gptoss} show commentary tokens which are interpretable for the end user, and are admittedly distinct from the actual intermediate tokens. The commentary tokens might very well have been generated post-hoc to explain the LLM's predicted answer. These are the different kinds of explanations - reasoning traces, their summaries or post hoc explanations -  which users are exposed along with the predicted answer when they interact with these systems.



With the widespread adoption of LLMs, especially in safety critical domains, the need to study the science of appropriate trust - trusting correct answers and distrusting incorrect answers - when interacting with a noisy system is paramount. Typical studies are often anecdotal and fail to disentangle the connection between the user's ability to verify the solution  and effectiveness of the provided explanation~\citep{fok}. If the user can independently verify the correctness of the predicted answer--either through their own knowledge or with access to external tools, they are not dependent on the explanation to aid them and thus, they often introduce bias in evaluating the explanations, may even choosing to ignore them. In such studies~\citep{bucina}, it is extremely difficult to convincingly isolate the effects of the provided explanations on user trust.

To address this gap, we propose an evaluation protocol for measuring how LLM-generated explanations affect user trust. The protocol simulates a setting where users do not have the ability or means to independently verify the LLM generated answer. For each problem, we expose users to different types of explanations alongside AI-generated answers and the participants need to judge whether the AI's predicted answer is correct or incorrect. We instantiate this protocol with a between-subjects user study on challenging math, physics, and chemistry problems from the JEE-Bench dataset~\citep{jee}. 

Our study reveals that reasoning traces and post-hoc explanations significantly increased false trust engendered in users, as compared to our baseline where no explanation was provided. In an attempt to mitigate this critical issue, we evaluate a post hoc explanation generation paradigm - dual explanations, similar to contrastive explanation generation in \citep{si-etal-2024-large}, where the user is exposed to the merits and demerits of the LLM generated answer. Overall, we find that the dual explanations have the lowest rate of engendering false trust. We also observe that reasoning traces and dual explanations achieve the highest rates of appropriate trust, though for fundamentally different reasons. Reasoning traces tend to increase user confidence in LLM-generated answers overall, leading to high accuracy in identifying correct answers but poor performance in detecting incorrect ones. In contrast, dual explanations exhibits a more balanced effect: users maintain relatively high accuracy in identifying both correct and incorrect answers. 

Our protocol also enables us to evaluate how the familiarity of the user with a problem instance changes the engendered trust rates. We found that when a user is familiar with a problem instance, reasoning trace or their summaries seem to be most helpful but when users are not confident in their ability to solve the problem, dual explanations helps the most and has the least false trust rate. 

Finally, we also observe that users do not find all explanations equally helpful in understanding how the LLM arrived at its predicted answer. Surprisingly, the supposed provenance explanations - reasoning traces and their summaries - were perceived as the least helpful for explaining the model's reasoning process.

In summary, we propose an evaluation protocol to isolate the effects of different explanation conditions on user trust. Through a carefully conducted between-subject user study, we found that reasoning traces, their summaries and post hoc LLM generated answer explanations can hinder the detection of incorrect AI answers and engender false trust in users. 
Finally, we find that \textit{dual explanations}, a post-hoc contrastive explanation generation method which presents both merits and demerits of the AI's answer, minimizes false trust while maintaining high detection accuracy.

%% file: sections/related_works.tex
\section{Related works}
\textbf{LLM Uncertainty and confidence measurement:}
\citet{steyver} study the gap between user confidence in LLM answers and the actual confidence of the model, finding that users overestimate LLM accuracy when provided with default explanations.~\citet{xiong} evaluate methods for eliciting verbalized confidence from LLMs and find that models are often overconfident, particularly on tasks requiring specialized knowledge. 
These work measures whether models can accurately express their uncertainty in their answers, and directly motivates our evaluation protocol for measuring how this affects user's trust.

\textbf{Hallucinations in LLMs:}
LLMs are known to produce fluent, confident outputs that are factually incorrect, a phenomenon broadly termed hallucination~\citep{ji, selfcheck}. Hallucinations have been documented across high-stakes domains such as law~\citep{Dahl_2024}. Hallucinations in such domain have real-world consequences for the user interacting with them. The danger is amplified by the fact that hallucinated outputs are often indistinguishable in tone and fluency from correct ones, making them difficult for non-expert users to detect. Our work addresses the question: when LLMs hallucinate, do the explanations provided alongside their answers help users catch the error, or do they make users more likely to accept it? 

\textbf{User Studies with LLM Explanations:}
~\citet{kim} find that LLM explanations increase user reliance on both correct and incorrect answers, while sources and inconsistencies reduce reliance on incorrect answers. Their work studies whether explanations should be present at all in a setting with general knowledge questions. We ask: given that explanations are ubiquitous, which \textit{types} of explanations engender false trust? ~\citet{sharma} found that users rate LLM responses with unfaithful explanations as equally trustworthy as those with genuine explanations when shown in isolation, suggesting that self-reported trust does not distinguish the quality of the explanation.~\citet{bo} evaluate different kinds of interventions for reliance on LLM advice and find that while interventions reduce over-reliance, they fail to improve appropriate reliance. In our work, we test reasoning traces from Large Reasoning Models, which these studies do not examine. Moreover, unlike these works, we test user interactions with LLMs in a setting where they cannot solve the problems independently in order to isolate the effect of the explanations on the user's judgment and trust.

\textbf{XAI Evaluation of Trust and Reliance:}
The xAI literature has shown that standard evaluation metrics for trust can be misleading ~\citep{inproceedings, schemmer, bansal, zhang, 10.1145/3287560.3287590, article}. \citet{bansal} demonstrate that explanations increase user agreement with AI recommendations regardless of whether the AI is correct. \citet{schemmer} propose metrics that separately measure user reliance on correct and incorrect AI advice, arguing that existing metrics like Weight on Advice do not capture whether users can distinguish good advice from bad. 
\citet{zhang} show that displaying confidence scores calibrates user trust but does not improve the accuracy of user decisions, as users cannot distinguish correct from incorrect outputs within the same confidence band.
 Together, these works establish that measuring trust alone is insufficient and what matters is whether that trust is warranted. 
 However, these frameworks were developed for traditional AI systems with structured explanations such as saliency maps and feature importance scores. Our evaluation protocol applies these insights to LLM-generated explanations, where the challenge is different: LLM explanations are expressed as fluent natural language which are not necessarily faithful to the process followed by the model to actually get to the answer.

%% file: sections/methods_2.tex
\section{Evaluation Protocol and Study Setup}
A key challenge in evaluating the effect of explanations on user trust is ensuring that the explanation is consequential. If users can independently verify the AI's answer, by solving the problem themselves without engaging with the explanations, they do not need the explanation at all. This would render any comparison across explanation types meaningless~\citep{fok}. Thus, we want carefully simulate a setting where users do not have the ability or means to verify the LLM's predicted answer but have enough background to meaningfully engage with the explanations.

\textbf{Study setup and Participant selection:}
To simulate the required setting, we conduct our evaluation on the JEE-Bench dataset~\citep{jee}, a challenging set of pre-engineering mathematics, physics, and chemistry problems from the IIT JEE-Advanced exam. Since these problems require significant domain expertise to solve, we recruited high school graduates via Prolific. These participants have foundational knowledge of physics, chemistry, and mathematics but have not yet acquired the domain expertise needed to solve competitive exam-level problems. This population represents the realistic use case for LLM explanations: users are qualified to engage with the provided explanations but not enough to verify the correctness of the predicted answer on their own. Users without sufficient domain knowledge may be unable to meaningfully interpret explanations grounded in subjects such as physics, mathematics, or chemistry. For them, every explanation is equally opaque.

\textbf{Simulated AI accuracy:}
The second design choice is controlling the LLM's accuracy in the study. We want to instantiate our protocol in a setting where LLMs are unreliable and it is very important that users actively decide when to trust the model. If the LLM's performance is near perfect, false trust is rare, since users who trust every answer are right most of the time, and all explanation conditions appear to perform well regardless of whether users are actually evaluating the AI's output.
We therefore curate a balanced set of instances: each participant sees 8 problems — 4 where the AI answered correctly and 4 where it answered incorrectly. 
Of the 8 items, 4 are \emph{fixed} questions kept constant across all participants and conditions. Since every participant in every condition sees these same 4 problems, any difference in user judgment on these items can be attributed directly to the explanation condition rather than to variation in problem difficulty. The remaining 4 are \emph{sampled} randomly from the full dataset (2 correct, 2 incorrect per participant), so that our findings are not dependent on a small hand-picked set of problems and generalize across a broader range of questions from the dataset.

\textbf{Explanation Conditions:}
We aim to cover the spectrum of explanation types that users encounter when interacting with LLMs. Each participant was randomly assigned to one of five conditions in a between-subjects design:
\begin{enumerate}
    \item \textbf{Solution only}: the question and LLM-generated answer, serving as the baseline;
\item \textbf{Reasoning trace}: the question, the full reasoning trace (raw intermediate tokens generated during the model's inference) from DeepSeek R1, and the predicted answer;
\item \textbf{Reasoning summary}: the question, a summary of the reasoning trace, and the answer. This reflects how frontier models (OpenAI, Google, Anthropic) currently present reasoning to users, who see summaries rather than full reasoning traces;
\item $\mathbf{E^{+}}$: the question, the answer, and a post-hoc explanation arguing for the answer's correctness. This represents the common paradigm where an explanation is generated after the answer to justify it;
\item $\mathbf{E^{+/-}}$: the question, the answer, and a post-hoc explanation presenting both arguments for and against the correctness of the answer. The rationale here is that the two explanations will provide a more critical perspective to the end user to help them develop a more appropriate level of trust.\footnote{The idea here is akin to asking conference reviewers to provide arguments for as well as against the acceptance of a paper, with the AC using those arguments to develop an appropriate level of trust on the reviewer's rating}. 
\end{enumerate}
We provide examples of these different explanations generated for a question in Appendix \ref{app:example}.

\textbf{Models and Explanation Generation:}
We used DeepSeek R1~\citep{deepseek} to generate responses on JEE-Bench problems. DeepSeek R1 provides access to the model's intermediate tokens, unlike closed-source reasoning models, giving us actual reasoning traces alongside the final answer. We used GPT-4o-mini~\citep{gpt-4o-mini} to generate summaries of the reasoning traces. We used DeepSeek V3~\citep{v3} to generate post-hoc explanations on R1-generated answers. We provide the prompt templates used in each condition in the Appendix~\ref{app:prompt}.

\textbf{Procedure:}
The study proceeded in three phases. First, participants answered background questions about their experience with competitive science and math exams and their frequency of AI tool usage. Second, participants worked through 8 problems. For each problem, participants first rated their familiarity (``I know the answer immediately,'' ``I might be able to solve it given more time,'' or ``I cannot solve this''). Participants who claimed immediate knowledge provided their own answer before seeing the AI's output; others proceeded directly to the AI's answer and explanation. Participants then judged whether the AI's answer was correct, and rated their confidence, trust, and the explanation's helpfulness. Third, participants completed a post-study questionnaire rating their overall trust in the AI system. More details about the user study is provided in Appendix~\ref{app:user}.

\textbf{Incentives:}
Participants received a base pay of \$3.00 and could earn a performance bonus of up to \$2.00 based on the accuracy of their judgments (\$0.25 per correct judgment across 8 items), for a maximum total of \$5.00. This gives participants an actual stake in judging the AI answer correctly. Without this, participants may default to accepting or rejecting the AI's answer without engaging with the explanation, which would undermine the evaluation.

The study was reviewed and approved by the Institutional Review Board (IRB). Participants were randomly assigned to one condition, with 25 participants per condition. The expected study duration was approximately 20 minutes.

%% file: sections/results_v2.tex
\section{Results}

\subsection{Do variants of reasoning traces and post-hoc explanations engender false trust?}

For each question across every condition, we asked the participant whether they thought the AI's predicted answer was correct. We can compare the user's perceived correctness of the AI's answer with the ground truth. We define \textit{false trust} as the conditional probability that an answer deemed correct by the user is actually incorrect. 
\begin{equation}
\text{False Trust} = \frac{|\{\text{responses judged as correct} \cap \text{actually incorrect}\}|}{|\{\text{responses judged as correct}\}|}
\end{equation}

\begin{table}[h]
\centering
\begin{tabular}{lcc}
\toprule
Condition & Perceived Correct (\%) & False Trust (\%) \\
\midrule
$E^{+/-}$            & 57.3 & \textbf{40.9} \\
$E^{+}$            & 66.0 & 43.9 \\
Reasoning summary & 58.5 & 43.6 \\
Reasoning trace   & 64.5 & 41.9 \\
Solution only     & 57.3 & 47.3 \\
\bottomrule
\end{tabular}
\vspace{0.1em}
\caption{\emph{Perceived correct (\%)} is the percentage of AI answers users judged to be correct (regardless of actual correctness); \emph{False trust (\%)} rates per explanation condition. \textbf{Bold} = best in column.}
\label{correct_and_false}
\end{table}

As shown by Table \ref{correct_and_false}, users are being persuaded into believing a greater number of LLM generated answers as correct when exposed to reasoning traces and post hoc LLM generated explanations.
This would have been a positive result if the False Trust (\%) rates associated with these conditions were low. However, we see that the $E^{+}$ has a very high False trust rate compared to other conditions.\footnote{Solution only condtion has the highest false trust but we believe it is because the users have limited domain knowledge and thus are enable to distinguish between a correct and an incorrect answer when exposed only to the LLM generated answer.} So, the number of responses where the user was persuaded into believing an incorrect llm generated answer is very high in the $E^{+}$ condition. On the other hand, the $E^{+\-}$ condition is more calibrated. Since the AI's accuracy in the task is 50\%, the rate of responses judged as correct is closest to the ground truth in the $E^{+/-}$ condition. Among the responses being judged as correct, the $E^{+/-}$ condition has the least False trust rate and thus, exposing the user to the merits and demerits of an answer helps in reducing the false trust engendered in the user.
\subsection{Appropriate Trust: Can users discriminate between correct and incorrect AI answers?}
The previous section talked about false trust, which provide insights into user behavior: how often users' trust is misplaced. 
However, these rates alone do not tell us which explanation was the most helpful to distinguish between correct and incorrect AI answers for the user. A user who simply answered all questions at random will have 50\% responses judged as correct, while a user who simply judged every answer as wrong will have the lowest false trust. Neither is desirable, as we want the explanations to help users accept correct answers \textit{and} reject incorrect answers.
\begin{table}[t]
\centering
\caption{Confusion-matrix metrics per condition (pooled items).
Precision, recall, and accuracy are in percent; F1 and
F$_{0.5}$ are unitless. F$_{0.5}$ weights precision over recall
($\beta{=}0.5$). \textbf{Bold} = best in column.}
\vspace{0.15em}
\label{tab:h2-confusion}
\small
\begin{tabular}{l rrrrr}
\toprule
Condition & Precision & Recall & F1 & F$_{0.5}$ & Accuracy \\
\midrule
Solution only     & 53.3 & 60.9 & .569 & .547 & 53.8 \\
$E^{+}$            & 56.2 & 75.0 & .643 & .592 & 58.3 \\
$E^{+/-}$            & \textbf{59.1} & 67.7 & .631 & \textbf{.606} & \textbf{60.4} \\
Reasoning summary & 55.4 & 64.6 & .596 & .570 & 56.2 \\
Reasoning trace   & 57.9 & \textbf{76.0} & \textbf{.658} & .589 & \textbf{60.4} \\
\bottomrule
\end{tabular}
\end{table}
To assess which condition helped users \emph{distinguish} correct from incorrect AI answers, we frame each participant's judgment as a binary classification problem and report standard confusion matrix metrics (Table ~\ref{tab:h2-confusion}). 
\begin{itemize}
    \item \textbf{Accuracy} $= \frac{TP + TN}{TP + TN + FP + FN}$.
    Fraction of judgments where the user correctly identified whether the AI was right or wrong.
    \item \textbf{Precision} $= \frac{TP}{TP + FP}$.
    When a user judges the AI's answer to be correct, how often is the AI actually correct? Low precision means the user's endorsement is unreliable.
    \item \textbf{Recall} $= \frac{TP}{TP + FN}$.
    How many correct AI answers did the user successfully judge to be correct?
    \item \textbf{F1} $= \frac{2 \cdot \text{Precision} \cdot
    \text{Recall}}{\text{Precision} + \text{Recall}}$. Balances precision and recall, but does not penalize false trust directly.

    \item $\mathbf{F_{\beta}}$ $= (1+\beta^{2})\cdot \frac{\text{Precision} \cdot \text{Recall}}{\beta^{2}\cdot\text{Precision} + \text{Recall}}$.We report $F_{0.5}$ as being relevant to settings where a user judging an incorrect AI answer as correct is costlier than judging a correct AI answer as incorrect.
\end{itemize}
At first glance, Reasoning trace and $E^{+/-}$ appear to be the strongest conditions, as they have the same accuracy. However, they achieve this in fundamentally different ways. Reasoning trace and $E^{+}$ both achieve high recall by shifting users toward saying 
``correct'' more frequently, on both correct \emph{and} incorrect items. In contrast, $E^{+/-}$ is the only condition that reduces false trust below the Solution only baseline while raising earned trust above it, achieving the highest precision (59.1\%) and tying for the highest accuracy (60.4\%).

To further examine the differences in user behavior when exposed to different explanations, we condition on AI correctness and introduce two additional metrics that capture how effectively users distinguish between correct and incorrect AI answers. Specifically, we report \textit{correct judgment} — the rate at which users accept correct AI answers—and \textit{misjudgment} — the rate at which users accept incorrect AI answers—in Table~\ref{tab:discrimination}.
The conditions $E^{+}$ and Reasoning trace achieve the highest correct judgment rates (74.0\% and 75.0\%), which implies that users in these conditions are good at recognizing correct AI answers. However, their misjudgment rates are also high (58.0\% and 54.0\%), meaning that these explanations push users toward accepting AI answers indiscriminately, regardless of their correctness. 

\begin{table}[h]
\centering
\caption{Correct judgment, Misjudgment, and Accuracy per explanation condition. \textbf{Bold} = best in column.}
\vspace{0.15em}
\label{tab:discrimination}
\small
\begin{tabular}{l rrr}
\toprule
Condition & Correct judgment ($\uparrow$) & Misjudgment ($\downarrow$) & Accuracy ($\uparrow$) \\
\midrule
Solution only     & 60.4 & 54.2 & 53.8 \\
$E^{+}$            & 74.0 & 58.0 & 58.3 \\
$E^{+/-}$            & 67.7 & \textbf{46.9} & \textbf{60.4} \\
Reasoning summary & 66.0 & 51.0 & 56.2 \\
Reasoning trace   & \textbf{75.0} & 54.0 & \textbf{60.4} \\
\bottomrule
\end{tabular}
\end{table}

$E^{+/-}$ shows a different pattern. It achieves the lowest misjudgment of any condition (46.9\%) while maintaining a correct judgment rate (67.7\%) that is above the Solution only baseline (60.4\%). This is the only condition where users accept more correct answers \emph{and} reject more incorrect answers relative to Solution only. $E^{+/-}$ improves discrimination ability of the users rather than just shifting them towards trusting the AI answers.
\subsection{How does the Appropriate trust change across Explanation conditions with instance level familiarity?}

Even though we selected users who do not have domain expertise to across all the presented questions, there will be cases where the participants are familiar with specific questions. Thus to isolate the effects of explanations across different instance level familiarity, participants in the user study rated their familiarity with each problem (``I know the answer immediately,'' ``I might be
able to solve it given more time,'' or ``I cannot solve this''). Table ~\ref{table:know1} shows the accuracy and the false trust rates across different explanation conditions for users grouped by their knowledge of the problem. When the users know how to solve the problem and can confidently reach to the solution, the need for explanation reduces~\citep{fok}. Users who are familiar with the problem can confidently verify whether the given AI prediction is correct or not without requiring any assistance in the form of an explanation. This is reflected by the highest accuracy seen in the case of solution only condition. We also see that such users, when displayed only the summary of the reasoning trace, can achieve very high performance while minimizing false trust. 

In cases where the users are not confident in their ability to solve the problem but have some familiarity with the problem, $E^{+/-}$ are the most helpful for the users. $E^{+/-}$ enable the users to have the best performance while simultaneously minimizing false trust. In contrast to the previous section where users knew how to solve the instances, the summary of the reasoning trace is the most detrimental for users among the conditions where an explanation is provided. Thus, users are unsure in their ability to solve the problem seem to be most susceptible to accept an incorrect answer when a succinct provenance explanation is provided along with it.

When users are completely unaware of how to solve the problem and are entirely reliant on the AI system, we see that explanations are always helpful and users perform much better in the $E^{+/-}$, Reasoning summary, and Reasoning trace conditions compared to the Solution only condition.
The performance in the $E^{+/-}$ condition is similar to the reasoning summary condition, but users are less likely to accept incorrect answers in the $E^{+/-}$ condition. Thus, even for lay users, giving the merits and demerits of a response helps reduce the risk of the LLM convincing the users its correct when its actually wrong.
\begin{table}[h]
\centering
\small
\begin{tabular}{llcc}
\toprule
Familiarity Level & Condition & Accuracy (\%) & False Trust (\%) \\
\midrule
I know it & $E^{+/-}$            & 61.5 & 28.6 \\
          & $E^{+}$            & 76.9 & 33.3 \\
          & Reasoning summary & 79.2 & 21.4 \\
          & Reasoning trace   & 63.6 & 33.3 \\
          & Solution only     & 80.0 & 29.2 \\
\midrule
I might solve it but need time 
          & $E^{+/-}$            & 64.0 & 35.7 \\
          & $E^{+}$            & 61.0 & 37.3 \\
          & Reasoning summary & 51.8 & 44.9 \\
          & Reasoning trace   & 61.7 & 41.3 \\
          & Solution only     & 47.4 & 51.0 \\
\midrule
I do not know how to solve 
          & $E^{+/-}$            & 56.7 & 45.9 \\
          & $E^{+}$            & 51.2 & 53.6 \\
          & Reasoning summary & 57.0 & 48.1 \\
          & Reasoning trace   & 58.9 & 43.9 \\
          & Solution only     & 46.7 & 54.3 \\
\bottomrule
\end{tabular}
\vspace{0.15em}
\caption{Accuracy and false-trust rates by explanation variant, segregated by self-reported familiarity with the question.}
\label{table:know1}
\end{table}
\subsection{Do users find all explanations equally helpful?}
After each question, across all conditions, participants rated whether the explanation helped them understand how the AI reached its answer. Solution only provided no explanation and is omitted from this analysis.

Ironically, we find that Reasoning traces, supposedly the provenance explanations, were rated as least helpful with only 56.5\% of responses being rated as helpful, and the post-hoc $E^{+}$ explanations, which were generated completely independently, were rated as most helpful for understanding how the AI reached its answer. 

We further analyzed the users' responses based on whether they find an explanation to be helpful or not. When users find a reasoning trace or its summary to be helpful, they end up believing that the LLM reasoned correctly and thus its predicted answer must be correct. This is reflected in the high perceived correct rates. However, their prediction accuracy is does not reflect the same trend. The better the explanations become in convincing the user that the answer is correct, the worse the user becomes at detecting the incorrect AI's predictions, having higher false trust rates and lower prediction accuracies. Ideally, we want explanations that are perceived as helpful by the users and help them distinguish between correct and incorrect LLM predictions. The $E^{+/-}$ satisfies both of the requirements - it has a high helpful score (72.4\%) and among those responses, users have the highest accuracy and lowest false trust rates.


 A significant number of questions (36\% and 43.6\%) have reasoning traces and their summaries that are perceived as not helpful by the users. We find that when users do not find these types of explanations to be helpful, their trust in the LLM generated predictions decreases, as indicated by the low perceived correct rates. This engendered distrust, due to the poor explanation quality, ends up being beneficial for the users as the majority of LLM predictions associated with these explanations are incorrect. Thus, the higher prediction accuracy and lower false trust rates should not be directly interpreted as evidence of higher explanation quality.

\begin{table}[h]
\centering
\small
\setlength{\tabcolsep}{5pt} 
\begin{tabular}{llcccc}
\toprule
Helpfulness & Variant & Total (\%) & Perceived Correct (\%) & Accuracy (\%) & False Trust (\%) \\
\midrule
Helpful 
& $E^{+/-}$            & 72.4 & 59.7 & 64.0 & 36.1 \\
& $E^{+}$            & 83.0 & 69.3 & 59.6 & 42.6 \\
& Reasoning summary & 64.0 & 77.3 & 57.0 & 44.4 \\
& Reasoning trace   & 56.5 & 86.7 & 55.8 & 43.9 \\
\midrule
Not Helpful 
& $E^{+/-}$            & 27.6 & 50.9 & 50.9 & 55.6 \\
& $E^{+}$            & 17.0 & 50.0 & 50.0 & 52.9 \\
& Reasoning summary & 36.0 & 25.0 & 58.3 & 38.9 \\
& Reasoning trace   & 43.5 & 35.6 & 66.7 & 35.5 \\
\bottomrule
\end{tabular}
\vspace{0.15em}
\caption{Distribution of responses and behavioral metrics across variants by perceived helpfulness.}
\label{Tab:help}
\end{table}

%% file: sections/discussions_v2.tex
\section{Discussion}
\textbf{Utility of the evaluation protocol: }In this paper, we introduce an evaluation protocol for measuring how LLM-generated explanations affect user trust. Since LLM-generated answers provide no correctness guarantees, the onus of judging the correctness falls on the user. Most frontier models either expose the users to LLM generated reasoning traces \citep{deepseek, qwq32b, qwen2.5_2024}, or show what can be a summarized version of the trace or a post-hoc explanation, alongside their answers \citep{jaech2024openai, singh2025openai, comanici2025gemini}. 
While we know that reasoning traces improve the performance of the model \citep{nye2021show, wei2022chain, deepseek, muennighoff2025s1}, the effect of exposing these traces to end-users is largely unstudied, specifically how they effect the user's trust. Our protocol is designed to fill this gap. By carefully controlling the domain knowledge of the participants, our protocol's design enables us to disentangle multiple dimensions of explanations related to user trust. 
Through our analysis of appropriate trust, correct judgment, and misjudgment, we were able to distinguish explanations that are merely persuasive from those that genuinely help users discriminate between correct and incorrect AI answers. By measuring the user's familiarity with the question instance, we were able analyze how the effects of the same explanation conditions change depending on the instance level familiarity in users. Our study also enabled us to get insights into which explanations users found most helpful for understanding how the AI reached to its prediction. 
Crucially, because our evaluation explicitly measures these dimensions, we can analyze how these dimensions are connected. It is critical to understand the implications of exposing these explanations along these dimensions, given the broad and rapid adoption of LLMs by end-users. 

\textbf{Reasoning traces and their summaries can hinder the detection of incorrect AI answers: } Our evaluation reveals that reasoning traces and their summaries have high false trust and misjudgement rates. This means that a user, when exposed to a reasoning trace or its summary leading to an incorrect prediction, is more likely to believe that the answer is correct. 
We speculate this occurs because reasoning traces sound plausible to users and resemble the kind of step-by-step thinking a human expert might produce while solving the problem. Users see familiar problem-solving language and phrases, and infer that since the model appears to be reasoning, the answer it arrived at must also be correct. 
This tendency is compelling enough that even researchers, based on qualitative analyses of trace contents, have hypothesized that reasoning traces improve LLM performance by instilling cognitive behaviors \citep{gandhi2025cognitive}. However, recent work~\citep{bs, int, kambhampati2026stop} shows that there is only a loose correlation between the correctness of the trace and the answer correctness. 
Moreover, authors in \citep{int} find that reasoning traces are least interpretable to users.
In \citep{anth, arcuschin2025chain}, researchers have also shown that the LLMs are not always faithful to their reasoning traces. Therefore, the disconnect between seemingly plausible reasoning traces and their summaries, and actual answer correctness is precisely what makes reasoning traces dangerous, as they engender false trust in users. 

\textbf{Post-hoc explanations can be overly persuasive and engender false trust in users: } Post-hoc explanations arguing for the answer's correctness also engenders high false trust and critically hamper the user's ability to distinguish between correct and incorrect answers. We speculate this is because these explanations are generated by models heavily optimized through RLHF to produce responses that satisfy users~\citep{rlhf}. In \citep{sree}, authors identify satisfaction as a key property of explanations in human-AI interaction : the explanation should leave the user feeling that they understand the AI's reasoning. LLM generated explanations excel at this, as they are optimized to produce responses that are helpful, warm, and satisfying to users. Recent work has shown that such optimization can increase sycophancy and reduce the accuracy of the model i.e. the tendency to produce responses that agree with or please the user rather than responses that are correct~\citep{ibrahim, chawla2026llms}. Models generating post-hoc explanations may be exhibiting similar tendencies.
 
\textbf{Dual side explanations help users to calibrate their trust:}
Motivated by the observation that one-sided explanations consistently increase false trust, we tested a dual explanation paradigm that presents both the merits and demerits of the AI's answer. The intuition is analogous to a peer review process: each reviewer presents both strengths and weaknesses of a submission, and the area chair makes the final decision by weighing both sides. Similarly, a dual explanations gives the user competing arguments and asks them to judge whether the answer is correct or incorrect. Our results show that this explanation is the only condition that reduces false trust below the no-explanation baseline while maintaining accuracy. 
We note that dual explanations are a first step and better explanation paradigms may further mitigate false trust engendered in users.

\textbf{Limitations: } While behavioral and self-reported measures are common approaches in human subject studies, they do inherit some of the hard to control variations in the subject answers. For example, 
our evaluation protocol assumes that participants cannot independently verify the AI's answers with other external tools during the study. While we selected JEE-Bench problems specifically because they require advanced domain expertise, we cannot fully guarantee that no participant had the knowledge to solve a given problem. 

%% file: sections/appendix.tex
\newpage
\appendix

\section{Prompt Templates}
\label{app:prompt}
\begin{prompttemplate}{Summarization Template: Reasoning Trace}
\ttfamily\small
Summarize the following reasoning trace in a very concise and clear manner, highlighting key events and outcomes in less than 250 words:\\[4pt]
\ldots\\
\{trace\}\\
\ldots\\[4pt]
Summary:
\end{prompttemplate}

\begin{prompttemplate}{Explanation Template: Correct Answer}
\ttfamily\small
You are given a \{subject\} problem and a predicted answer. Your task is to explain why this predicted answer is correct.\\[4pt]
Question: \{question\}\\[2pt]
Predicted Answer: \{answer\}\\[4pt]
Provide a concise explanation within 250 words for why this predicted answer is correct. Do not include any other text in your response.
\end{prompttemplate}

\begin{prompttemplate}{Explanation Template: Dual (Correct \& Incorrect)}
\ttfamily\small
You are given a \{subject\} problem and an AI-generated answer. Your task is to generate two sets of structured reasoning:\\[4pt]
1. \textbf{Reasons why the answer might be correct}\\
2. \textbf{Reasons why the answer might be incorrect}\\[4pt]
\textbf{Question:}\\
\{question\}\\[4pt]
\textbf{AI Answer:}\\
\{answer\}\\[4pt]
\textbf{Reasons why this answer might be correct (within 250 words):}\\[4pt]
\textbf{Reasons why this answer might be incorrect (within 250 words):}\\[4pt]
Do not include any other text outside of the format above.
\end{prompttemplate}

\begin{prompttemplate}{Solver Template (unified across answer types)}
\ttfamily\small
Solve the following \{subject\} problem and provide the answer in JSON format.\\[4pt]
QUESTION:\\
\{question\}\\[4pt]
\textnormal{\textit{[OPTIONS block, included only for MCQ variants:]}}\\
OPTIONS:\\
\{options\}\\[4pt]
\textnormal{\textit{[Answer-type instruction — see Table~\ref{tab:solver-variants}:]}}\\
\{answer\_instruction\}\\[4pt]
Provide your final answer in the following JSON format:\\[2pt]
\{\\
\hspace*{1em}"answer": "\{answer\_format\}"\\
\}
\end{prompttemplate}

\begin{table}[H]
\centering
\small
\begin{tabular}{@{}lll@{}}
\toprule
\textbf{Question type} & \textbf{Answer instruction} & \textbf{JSON value} \\
\midrule
MCQ (single) & Choose the single best answer. & \texttt{<A/B/C/D>} \\
MCQ (multiple) & One or more options may be correct. & \texttt{<letters>} \\
& Select all that apply. & \\
Integer & The answer is a non-negative integer. & \texttt{<integer>} \\
Numeric & The answer is a numerical value (may be & \texttt{<number>} \\
& decimal). Round to 2 decimal places if needed. & \\
\bottomrule
\end{tabular}
\caption{Variants of the solver prompt. The \texttt{OPTIONS} block is included only for MCQ variants.}
\label{tab:solver-variants}
\end{table}
%
%

%
%

\section{Example}
\label{app:example}
\paragraph{Question (shown to participants).}
The value of $\int_{-\frac{\pi}{2}}^{\frac{\pi}{2}} \frac{x^{2} \cos x}{1+e^{x}} d x$ is equal to

(A) $\frac{\pi^{2}}{4}-2$

(B) $\frac{\pi^{2}}{4}+2$

(C) $\pi^{2}-e^{\frac{\pi}{2}}$

(D) $\pi^{2}+e^{\frac{\pi}{2}}$

\paragraph{AI predicted answer.}
\textbf{A} \quad (gold answer: \textbf{A}; AI is \textbf{correct}).

\begin{tcolorbox}[breakable, enhanced,
  colback=gray!5, colframe=gray!50!black,
  title={(1) Reasoning trace },
  fonttitle=\bfseries\small, fontupper=\footnotesize]
\begin{Verbatim}[breaklines=true, breakanywhere=true, breaksymbolleft={}, breaksymbolright={}, fontsize=\footnotesize]
We are asked: "The value of $\int_{-\frac{\pi}{2}}^{\frac{\pi}{2}} \frac{x^{2} \cos x}{1+e^{x}} d x$ is equal to" with options (A), (B), (C), (D).

We need to evaluate the integral: I = integral_{-pi/2}^{pi/2} (x^2 cos x)/(1+e^x) dx.

This is a definite integral with symmetric limits and an integrand that involves an exponential factor. Often for integrals of the form integral_{-a}^{a} f(x)/(1+e^x) dx, we use the property that integral_{-a}^{a} f(x)/(1+e^x) dx = integral_{0}^{a} f(x) dx if f(x) is even? Actually, there is a known trick: For an integrand of the form f(x)/(1+e^x) over symmetric limits, we can combine with the substitution u = -x to get another integral and then add. More specifically, let I = integral_{-a}^{a} f(x)/(1+e^x) dx. Consider substitution x -> -x: I = integral_{-a}^{a} f(-x)/(1+e^{-x}) dx = integral_{-a}^{a} f(-x) e^x/(1+e^x) dx. If f is even, then f(-x)=f(x), so I = integral_{-a}^{a} f(x) e^x/(1+e^x) dx. Adding the original I and this one: 2I = integral_{-a}^{a} f(x) (1/(1+e^x) + e^x/(1+e^x)) dx = integral_{-a}^{a} f(x) dx. So if f is even, then I = (1/2) integral_{-a}^{a} f(x) dx = integral_{0}^{a} f(x) dx. That is a standard result: integral_{-a}^{a} f(x)/(1+e^x) dx = integral_{0}^{a} f(x) dx when f is even.

But is our f(x) even? Here the integrand is (x^2 cos x)/(1+e^x). The function f(x) = x^2 cos x is indeed even because x^2 is even, cos x is even. So we can apply that property. However, we should verify the conditions: The property holds if f is continuous and even, and the limits are symmetric. So we have:
I = integral_{-pi/2}^{pi/2} (x^2 cos x)/(1+e^x) dx = integral_{0}^{pi/2} x^2 cos x dx.
But wait: Is that property always valid? Let's check: 
Let I = integral_{-a}^{a} f(x)/(1+e^x) dx, where f is even. Set u = -x, then I = integral_{a}^{-a} f(-u)/(1+e^{-u}) (-du) = integral_{-a}^{a} f(u)/(1+e^{-u}) du = integral_{-a}^{a} f(u) e^u/(1+e^u) du. So I = integral_{-a}^{a} f(x) e^x/(1+e^x) dx. Adding the two expressions: I + I = integral_{-a}^{a} f(x)[1/(1+e^x) + e^x/(1+e^x)] dx = integral_{-a}^{a} f(x) dx. So indeed 2I = integral_{-a}^{a} f(x) dx, so I = (1/2) integral_{-a}^{a} f(x) dx = integral_{0}^{a} f(x) dx. Yes, it's valid provided the integrals converge. So we get:
I = integral_{0}^{pi/2} x^2 cos x dx.

Now we need to compute J = integral_{0}^{pi/2} x^2 cos x dx.

We can compute this using integration by parts. Let u = x^2, dv = cos x dx, then du = 2x dx, v = sin x. So:
integral x^2 cos x dx = x^2 sin x - integral 2x sin x dx = x^2 sin x - 2 integral x sin x dx.
Now compute integral x sin x dx: u = x, dv = sin x dx, du = dx, v = -cos x, so integral x sin x dx = -x cos x + integral cos x dx = -x cos x + sin x + C.
Thus, integral x^2 cos x dx = x^2 sin x - 2(-x cos x + sin x) + C = x^2 sin x + 2x cos x - 2 sin x + C = (x^2 - 2) sin x + 2x cos x + C.

Now evaluate from 0 to pi/2:
At x = pi/2: sin(pi/2)=1, cos(pi/2)=0, so ( (pi/2)^2 - 2 ) * 1 + 2*(pi/2)*0 = (pi^2/4 - 2).
At x = 0: sin(0)=0, cos(0)=1, so (0^2 - 2)*0 + 2*0*1 = 0.
Thus, J = (pi^2/4 - 2) - 0 = pi^2/4 - 2.

So I = pi^2/4 - 2.

Now look at the options:
(A) pi^2/4 - 2
(B) pi^2/4 + 2
(C) pi^2 - e^(pi/2)
(D) pi^2 + e^(pi/2)

So the answer is (A).

But we must be cautious: The property we used is valid only if f is even. f(x)=x^2 cos x is indeed even. But also the integration limits are symmetric. So it seems correct.

We can also verify by direct computation or consider if there is any issue with the integrand at some point? It is continuous. So answer should be (A).

Thus, final answer is A.

We need to provide in JSON: {"answer": "A"}

But wait: The options are labeled (A), (B), (C), (D). So we put "A" in the string.

Thus final output: { "answer": "A" }
\end{Verbatim}
\end{tcolorbox}

\begin{tcolorbox}[breakable, enhanced,
  colback=blue!3, colframe=blue!40!black,
  title={(2) Reasoning-trace summary (condensed by a separate model)},
  fonttitle=\bfseries\small, fontupper=\small]
The problem involves evaluating the integral \( I = \int_{-\frac{\pi}{2}}^{\frac{\pi}{2}} \frac{x^2 \cos x}{1+e^x} \, dx \). The integrand is symmetric, and since \( f(x) = x^2 \cos x \) is even, we can apply the property that \( \int_{-a}^{a} \frac{f(x)}{1+e^x} \, dx = \int_{0}^{a} f(x) \, dx \). This simplifies our integral to \( I = \int_{0}^{\frac{\pi}{2}} x^2 \cos x \, dx \).

Next, we compute \( J = \int_{0}^{\frac{\pi}{2}} x^2 \cos x \, dx \) using integration by parts. After performing the necessary calculations, we find \( J = \frac{\pi^2}{4} - 2 \). Thus, \( I = J = \frac{\pi^2}{4} - 2 \).

The answer options are (A) \( \frac{\pi^2}{4} - 2 \), (B) \( \dots \) 
\end{tcolorbox}

\begin{tcolorbox}[breakable, enhanced,
  colback=green!4, colframe=green!40!black,
  title={(3) $E_{+}$ : Explanation supporting the predicted answer},
  fonttitle=\bfseries\small, fontupper=\small]
The integral is of the form \( I = \int_{-a}^{a} \frac{x^2 \cos x}{1+e^x} \, dx \).  
Using the property \( \int_{-a}^{a} f(x) \, dx = \int_{-a}^{a} f(-x) \, dx \), we note that  
\[
f(-x) = \frac{(-x)^2 \cos(-x)}{1+e^{-x}} = \frac{x^2 \cos x}{1+e^{-x}}.
\]  
Adding \( f(x) \) and \( f(-x) \):  
\[
f(x) + f(-x) = x^2 \cos x \left( \frac{1}{1+e^x} + \frac{1}{1+e^{-x}} \right).
\]  
Since \( \frac{1}{1+e^{-x}} = \frac{e^x}{1+e^x} \), the sum inside the parentheses is \( 1 \).  
Thus \( f(x) + f(-x) = x^2 \cos x \).  

Therefore,  
\[
I = \frac12 \int_{-\pi/2}^{\pi/2} [f(x) + f(-x)] \, dx = \frac12 \int_{-\pi/2}^{\pi/2} x^2 \cos x \, dx.
\]  

The integrand \( x^2 \cos x \) is even, so  
\[
I = \int_{0}^{\pi/2} x^2 \cos x \, dx.
\]  

Integrate by parts:  
Let \( u = x^2 \), \( dv = \cos x \, dx \), then \( du = 2x \, dx \), \( v = \sin x \).  
First step:  
\[
\int_{0}^{\pi/2} x^2 \cos x \, dx = \left[ x^2 \sin x \right]_{0}^{\pi/2} - \int_{0}^{\pi/2} 2x \sin x \, dx.
\]  
The boundary term: \( (\pi^2/4) \cdot 1 - 0 = \pi^2/4 \).  

Now \( \int_{0}^{\pi/2} 2x \sin x \, dx \):  
Integrate by parts again: \( u = 2x \), \( dv = \sin x \, dx \), \( du = 2 dx \), \( v = -\cos x \).  
\[
\int_{0}^{\pi/2} 2x \sin x \, dx = \left[ -2x \cos x \right]_{0}^{\pi/2} + \int_{0}^{\pi/2} 2 \cos x \, dx.
\]  
Boundary: \( -2 \cdot (\pi/2) \cdot 0 + 0 = 0 \) (since \( \cos(\pi/2) = 0 \)), plus \( \int_{0}^{\pi/2} 2 \cos x \, dx = [2\sin x]_{0}^{\pi/2} = 2 \).  

Thus:  
\[
I = \frac{\pi^2}{4} - 2.
\]  

This matches option (A).
\end{tcolorbox}

\begin{tcolorbox}[breakable, enhanced,
  colback=orange!5, colframe=orange!50!black,
  title={(4) $E_{\pm}$ (\texttt{e\_dual}) : Paired pro/con arguments},
  fonttitle=\bfseries\small, fontupper=\footnotesize]
\textbf{Reasons why this answer might be correct:}
\begin{Verbatim}[breaklines=true, breakanywhere=true, breaksymbolleft={}, breaksymbolright={}, fontsize=\footnotesize]
The integral simplifies using a symmetry property for integrals of the form integral_{-a}^{a} f(x)/(1+e^x) dx. Since f(x) = x^2 cos x is even, substitute x -> -x and add the original to obtain 2I = integral_{-pi/2}^{pi/2} x^2 cos x dx. Thus, I = (1/2) integral_{-pi/2}^{pi/2} x^2 cos x dx = integral_{0}^{pi/2} x^2 cos x dx due to evenness. Evaluating this via integration by parts: let u = x^2, dv = cos x dx, yielding [x^2 sin x] from 0 to pi/2 - integral 2x sin x dx = pi^2/4 - 2integral x sin x dx. Then integral x sin x dx from 0 to pi/2 gives [-x cos x + sin x] from 0 to pi/2 = 1. So I = pi^2/4 - 2, matching option A. This method is standard and avoids complex exponentials, relying on reliable calculus techniques.
\end{Verbatim}
\textbf{Reasons why this answer might be incorrect:}
\begin{Verbatim}[breaklines=true, breakanywhere=true, breaksymbolleft={}, breaksymbolright={}, fontsize=\footnotesize]
The symmetry property might be misapplied; if one incorrectly assumes f(x) is odd or mishandles the substitution, the reduction to (1/2)integral f(x) dx could fail. Errors in integration by parts for integral x^2 cos x dx could alter the result: if integral x sin x dx from 0 to pi/2 is computed as -1 instead of 1, the answer becomes pi^2/4 + 2 (option B). Forgetting the 1/2 factor after using symmetry would give 2(pi^2/4 - 2) = pi^2/2 - 4, not an option. Direct integration without symmetry might involve messy steps with e^x, leading to incorrect simplifications that produce options like pi^2 +/- e^(pi/2) (C or D). Additionally, overlooking that the integrand is not directly integrable in elementary terms could lead to approximate or symbolic errors, especially if software or AI miscalculates the exponential component.
\end{Verbatim}
\end{tcolorbox}
\section{User-study}
\label{app:user}
We conducted user-studies across 5 different conditions to evaluate the engendered user-trust across various paradigms.
Participants were recruited via Prolific and the study was
reviewed and approved by the Institutional Review Board (IRB code: SS64726).
In each condition, participants were shown 8 problems drawn from physics, chemistry, and mathematics. The AI system's accuracy on the selected items was 50\% (4 correct and 4 incorrect answers per participant). Of the 8 items, 4 were \emph{fixed} questions that were kept constant across all participants and conditions. The remaining 4 were \emph{sampled} questions, randomly drawn from the full dataset of correctly and incorrectly answered questions, with 2 correct and 2 incorrect answers per participant. 

Each participant was randomly assigned to only one of five conditions, making it a between-subjects study. In each condition, the participant saw only one of the following:
\begin{enumerate}
    \item \textbf{Solution-only: }The question and LLM-generated answer.
    \item \textbf{Reasoning trace: }The question, reasoning trace, and LLM-generated answer.
    \item \textbf{Summary: }The question, summary of the reasoning trace, and LLM-generated answer.
    \item \textbf{$E^{+}:$ }The question, LLM-generated answer, and an explanation of why the generated answer is correct.
    \item \textbf{$E^{+}$ and $E^{-}:$} The question, LLM-generated answer, and an explanation of why the generated answer might be correct and why the answer might be incorrect.
\end{enumerate}
\paragraph{Study Description:}
The study proceeded in three phases. First, participants answered background questions about their experience with competitive science/math exams and their frequency of AI tool usage.
\begin{itemize}
    \item \emph{``Have you ever prepared for or taken JEE (Mains or Advanced) or similar competitive science/math exams?''}
    \item \emph{``How often do you use AI tools (ChatGPT, Copilot, Claude, etc.)?''}
\end{itemize}
 
Second, participants worked through 8 problems. For each problem: 
\begin{enumerate}
    \item The participant read the problem statement and answer choices.
 
    \item The participant was asked: \emph{``Do you know how to solve this question?''} with three options: ``I know the answer immediately,'' ``I might be able to solve it given more time,'' or ``I cannot solve this.''
 
    \item \textbf{If the participant selected ``I know the answer immediately,''} they were shown the problem again and asked to select their own answer and rate their confidence: \emph{``How confident are you in your own answer?''} (7-point Likert scale). They were then shown the AI system's predicted answer along with the explanation corresponding to their assigned condition.
 
    \textbf{If the participant selected either of the other two options,} they were taken directly to the AI system's predicted answer and the explanation corresponding to their condition. In the \texttt{solution\_only} condition, no explanation was shown.
 
    \item The participant was asked: \emph{``Do you think the AI's answer is correct?''} (Yes/No). Participants were reminded that an incorrect judgment would cost \$0.25 from their bonus.
 
    \item The participant rated: \emph{``How confident are you in your judgment about the AI's answer?''} (7-point Likert scale).
 
    \item The participant rated: \emph{``I trust the AI system's outputs.''} (7-point Likert scale).
 
    \item The participant was asked: \emph{``Did the AI reasoning help you understand how the model reached its answer?''} (Yes/No). This question was omitted in the \texttt{solution\_only} condition.
\end{enumerate}

Third, participants completed a post-study questionnaire in which they rated the following statements on a 7-point Likert scale:
\begin{itemize}
    \item \emph{``I trust the AI system's outputs.''}
    \item \emph{``I would rely on the AI system's recommendations.''}
    \item \emph{``I am confident in the AI system's answers.''}
\end{itemize}

Participants received a base pay of \$3.00 and could earn a performance bonus of up to \$2.00 based on the accuracy of their judgments of the AI's answers (\$0.25 per correct judgment across 8 items), for a maximum total of \$5.00. This incentive structure was designed to encourage careful evaluation of the AI's outputs. The expected study duration was approximately 20 minutes.

\begin{figure}[ht!]
  \centering
  \fbox{\includegraphics[width=0.85\linewidth]{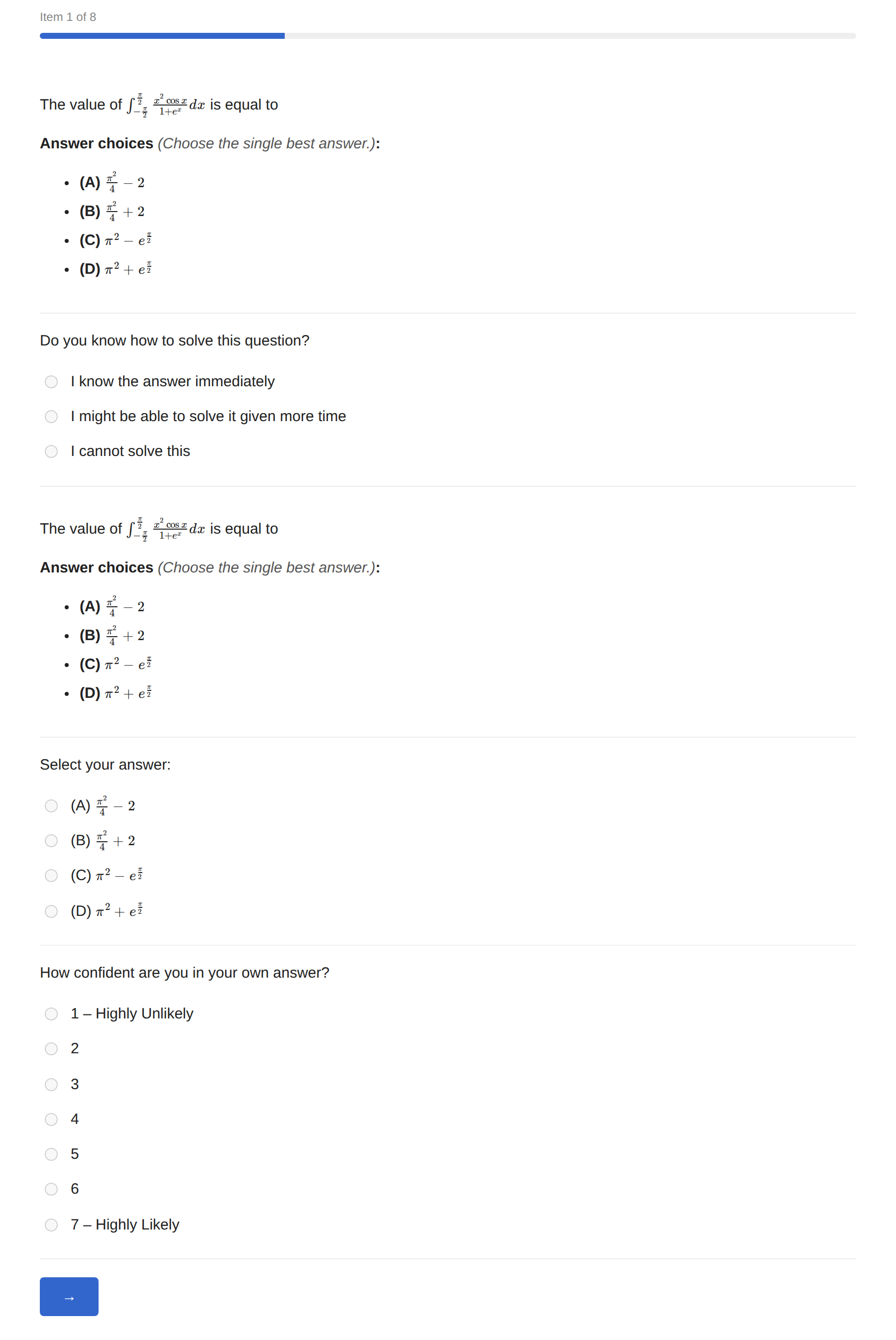}}
  \caption{\textbf{Step 5 -- Item page 1 (pre-AI).} The participant sees the
  problem statement, indicates whether they know how to solve it
  (\emph{know-level}: ``I know the answer immediately'' / ``given more time''
  / ``I cannot solve this''), optionally provides their own answer, and rates
  their confidence on a 1--7 Likert scale. This page is identical across
  variants.}
  \label{fig:flow-item-page1}
\end{figure}

\begin{figure}[ht!]
  \centering
  \fbox{\includegraphics[width=0.55\linewidth]{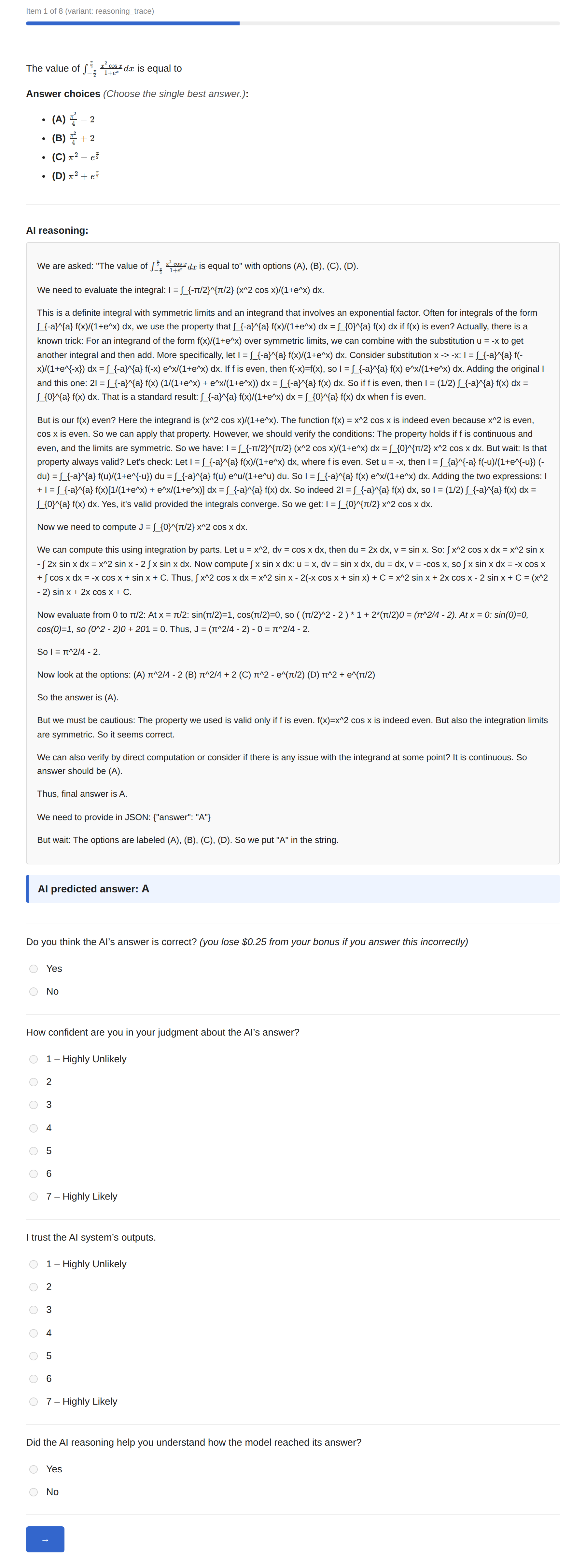}}
  \caption{\textbf{Step 6a -- Item page 2, \texttt{reasoning\_trace} variant.}
  After Page~1, the participant sees the AI's predicted answer alongside the
  full chain-of-thought trace. The page also collects the AI-correctness
  judgment (Yes/No), confidence in that judgment (1--7), per-item trust
  rating (1--7), and a helpfulness probe.}
  \label{fig:flow-trace}
\end{figure}

\begin{figure}[ht!]
  \centering
  \fbox{\includegraphics[width=0.85\linewidth]{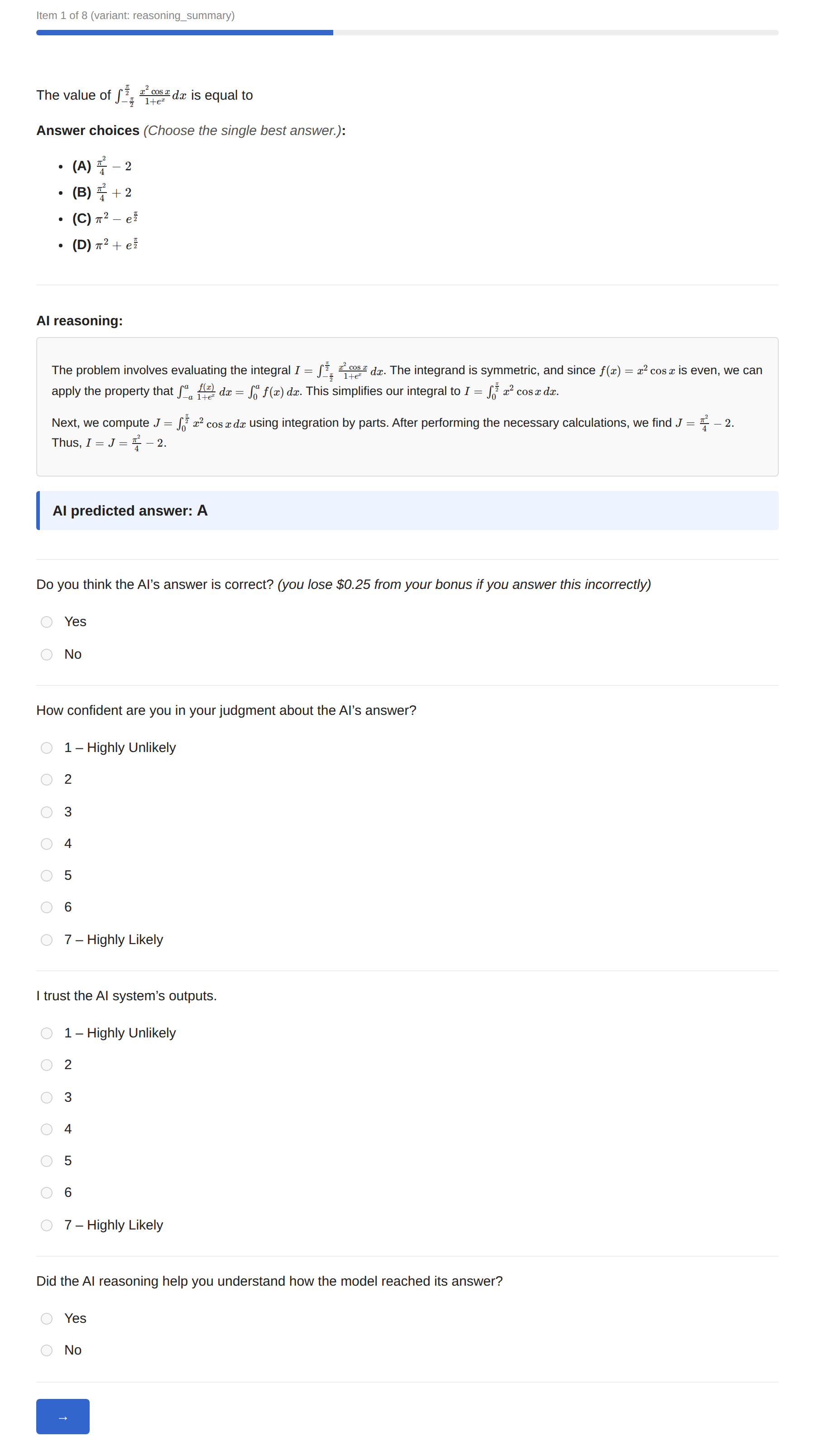}}
  \caption{\textbf{Step 6b -- Item page 2, \texttt{reasoning\_summary} variant.}
  Same layout as Figure~\ref{fig:flow-trace}, but the AI's reasoning is
  replaced by a condensed summary produced by a separate model.}
  \label{fig:flow-summary}
\end{figure}

\begin{figure}[ht!]
  \centering
  \fbox{\includegraphics[width=0.85\linewidth]{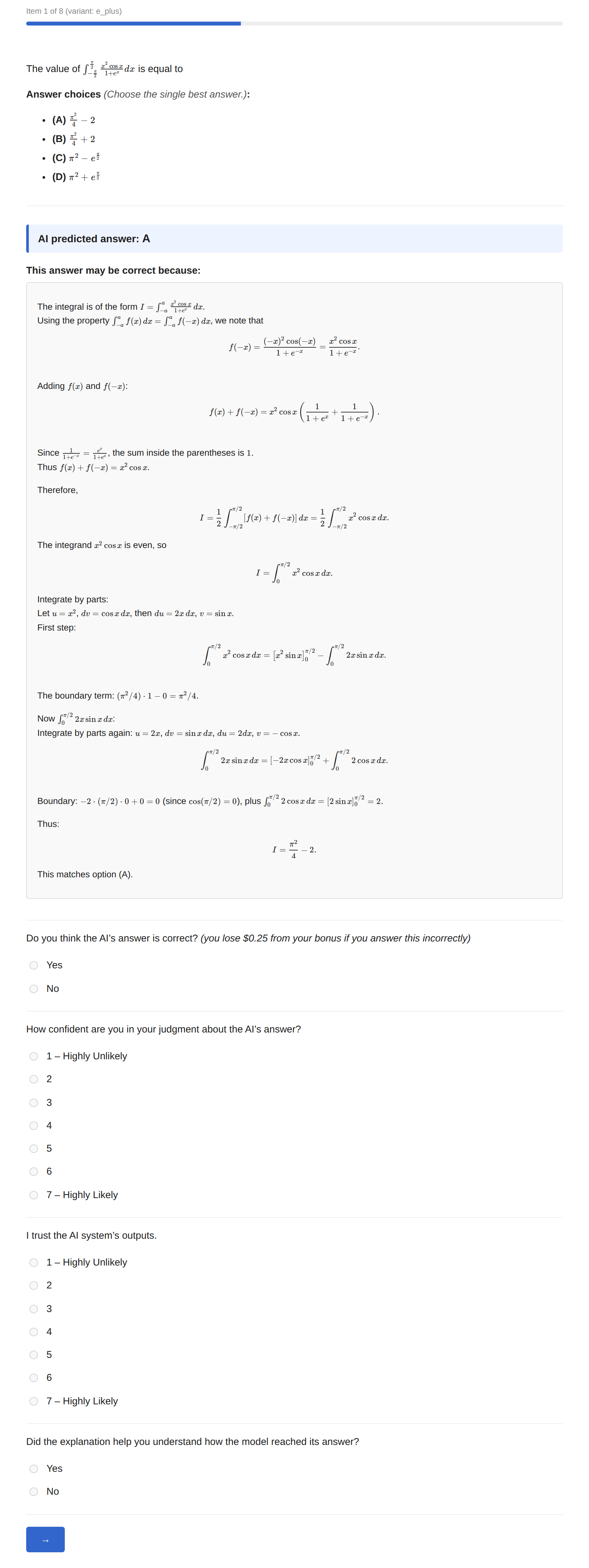}}
  \caption{\textbf{Step 6c -- Item page 2, $E_{+}$ variant.}
  Replaces the trace with a single-sided post-hoc explanation arguing
  \emph{why the predicted answer may be correct}.}
  \label{fig:flow-eplus}
\end{figure}

\begin{figure}[ht!]
  \centering
  \fbox{\includegraphics[width=0.80\linewidth]{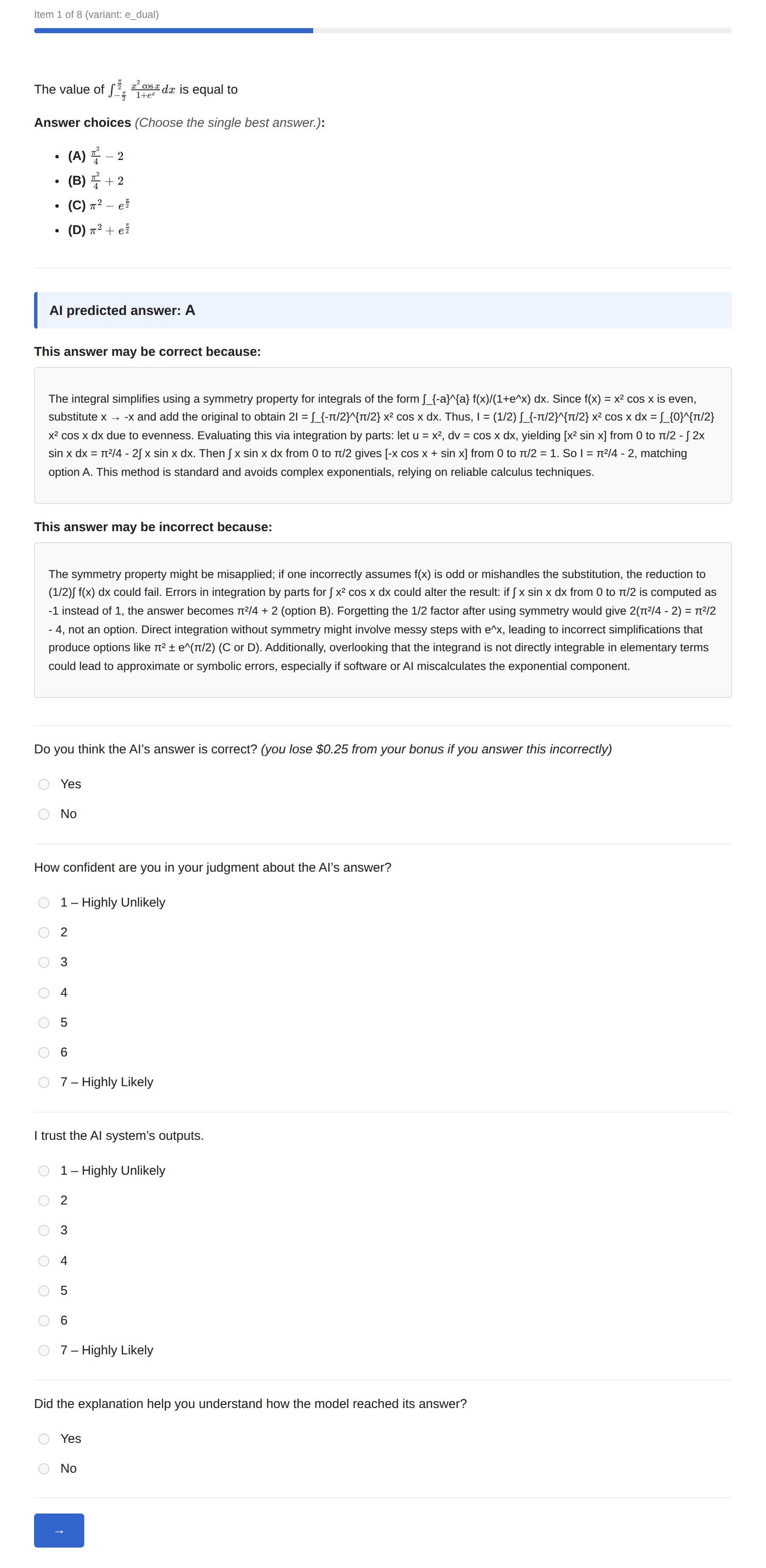}}
  \caption{\textbf{Step 6d -- Item page 2, $E^{+/-}$}
  Replaces the trace with paired ``why this answer may be correct'' /
  ``why this answer may be incorrect'' arguments.}
  \label{fig:flow-edual}
\end{figure}

\paragraph{Study Flow and example item walk-through}
This appendix reproduces the screens that participants saw, in order, during
the study. We illustrate the per-item flow on one fixed item
(idx~59, JEE-Bench Mathematics MCQ): the participant first sees the problem
and is asked whether they know how to solve it (Page~1; Figure~\ref{fig:flow-item-page1}),
and then sees the AI's answer along with the variant-specific explanation
(Page~2; Figures~\ref{fig:flow-trace}--\ref{fig:flow-edual}, one per variant).
Identical flow logic is used for the three other fixed items and for the
sampled correct/incorrect items in the per-participant pool.

For reference, the example item is:
\begin{quote}\small
The value of $\int_{-\frac{\pi}{2}}^{\frac{\pi}{2}} \frac{x^{2} \cos x}{1+e^{x}} d x$ is equal to

(A) $\frac{\pi^{2}}{4}-2$

(B) $\frac{\pi^{2}}{4}+2$

(C) $\pi^{2}-e^{\frac{\pi}{2}}$

(D) $\pi^{2}+e^{\frac{\pi}{2}}$
\end{quote}
The AI's predicted answer is \textbf{A} (gold answer:
\textbf{A}; the AI is \textbf{correct}).

\begin{figure}[ht!]
  \centering
  \fbox{\includegraphics[width=0.85\linewidth]{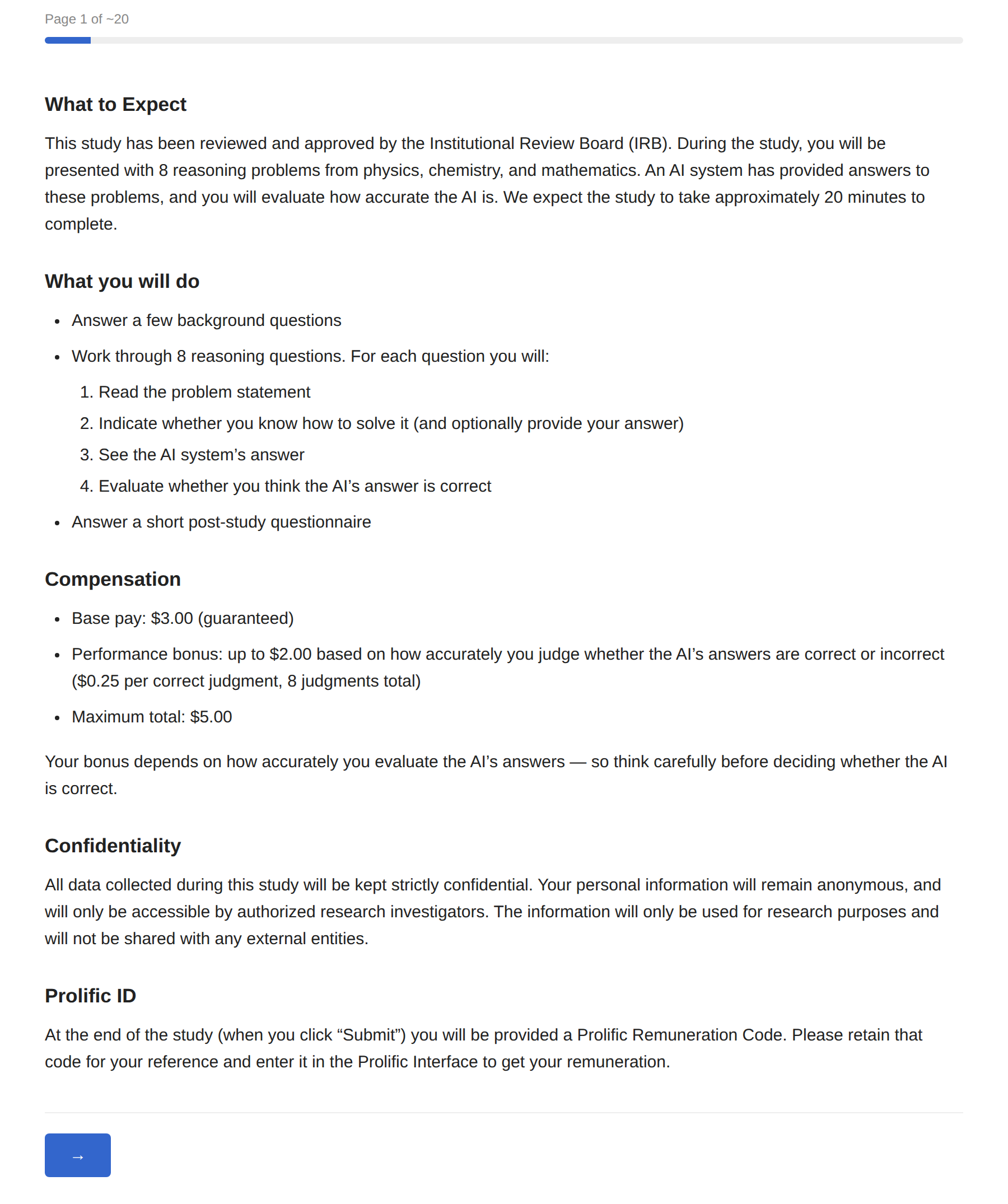}}
  \caption{\textbf{Step 1 -- Consent.} IRB-approved description of the study,
  expected duration, compensation structure (base pay plus 
  bonus), and Prolific completion procedure.}
  \label{fig:flow-consent}
\end{figure}
\begin{figure}[ht!]
  \centering
  \fbox{\includegraphics[width=0.92\linewidth]{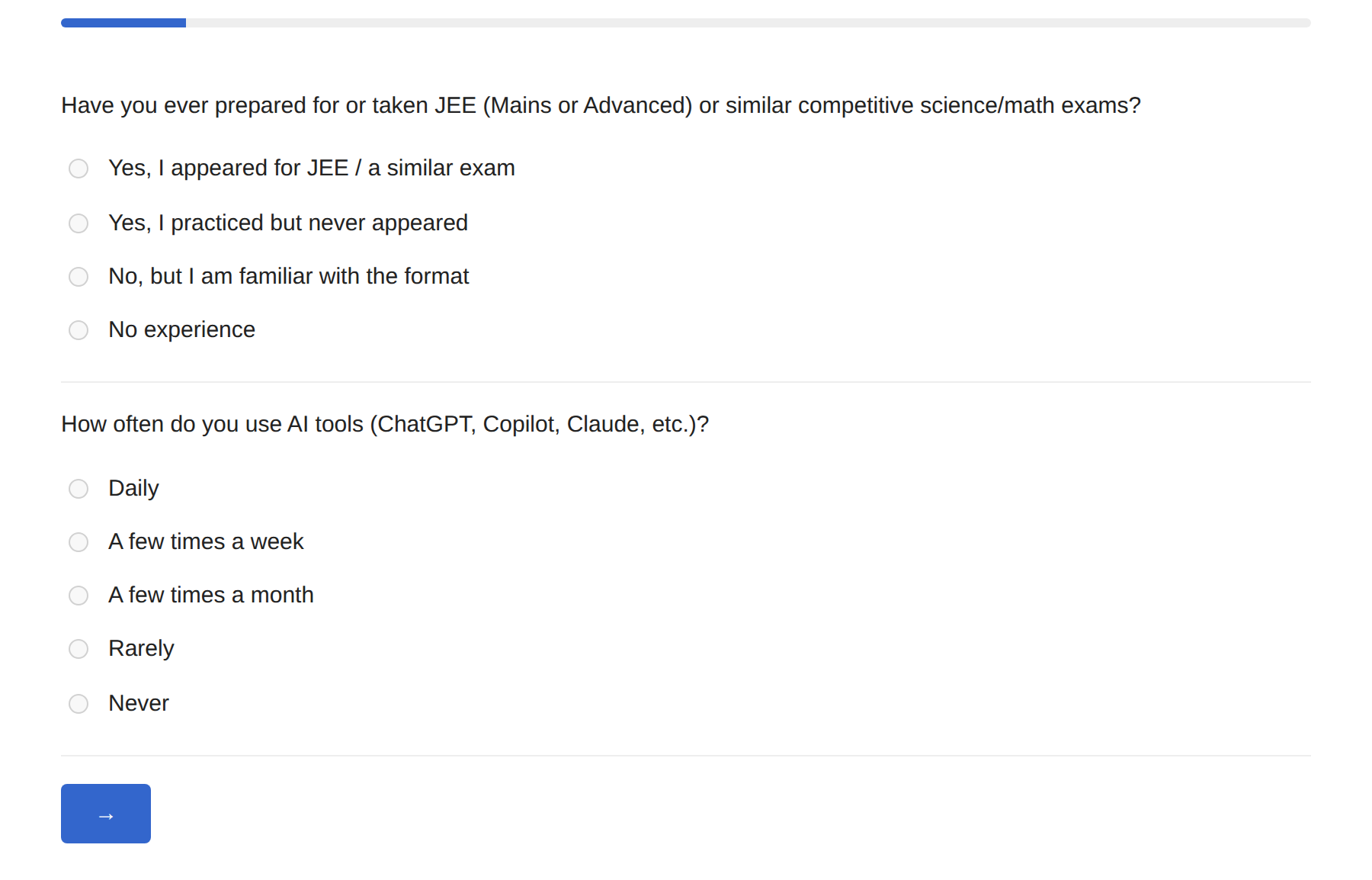}}
  \caption{\textbf{Step 2 -- Background questions.} Two single-choice items
  capture prior exposure to JEE-style competitive exams and frequency of
  AI-tool use. Used as covariates / for stratification.}
  \label{fig:flow-bg}
\end{figure}
\begin{figure}[ht!]
  \centering
  \fbox{\includegraphics[width=0.92\linewidth]{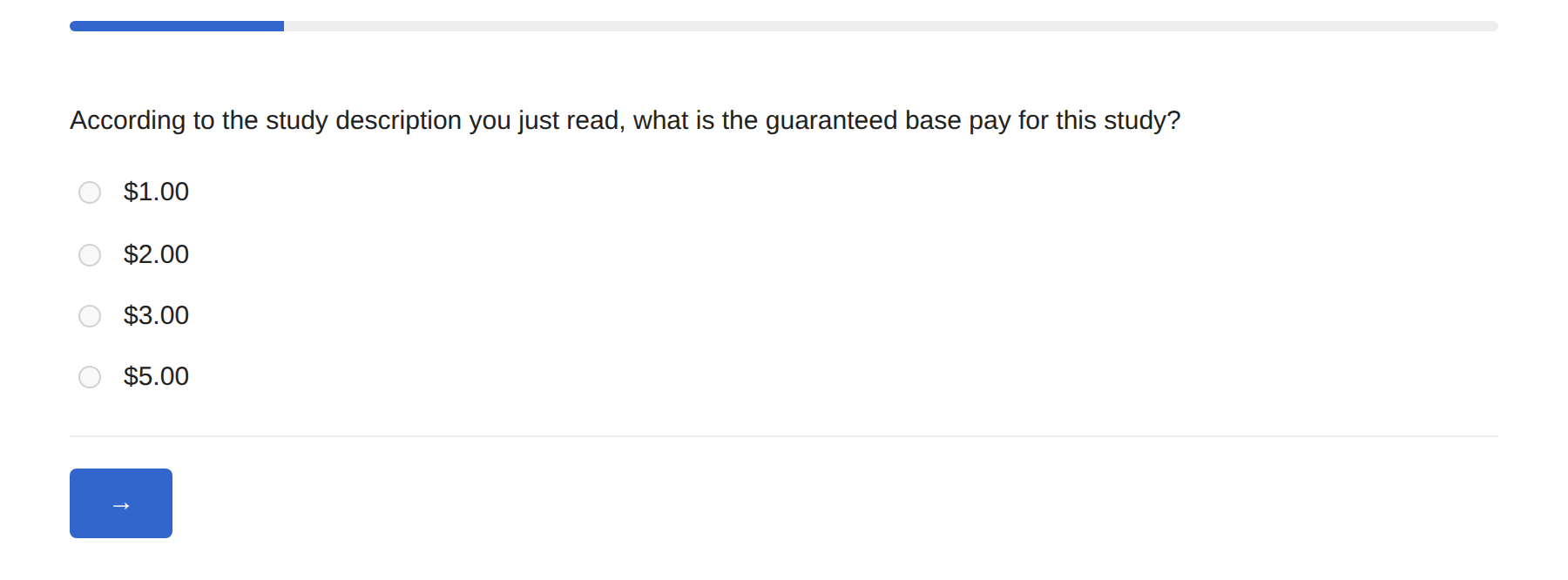}}
  \caption{\textbf{Step 3 -- Attention check 1.}}
  \label{fig:flow-attn1}
\end{figure}
\begin{figure}[ht!]
  \centering
  \fbox{\includegraphics[width=0.92\linewidth]{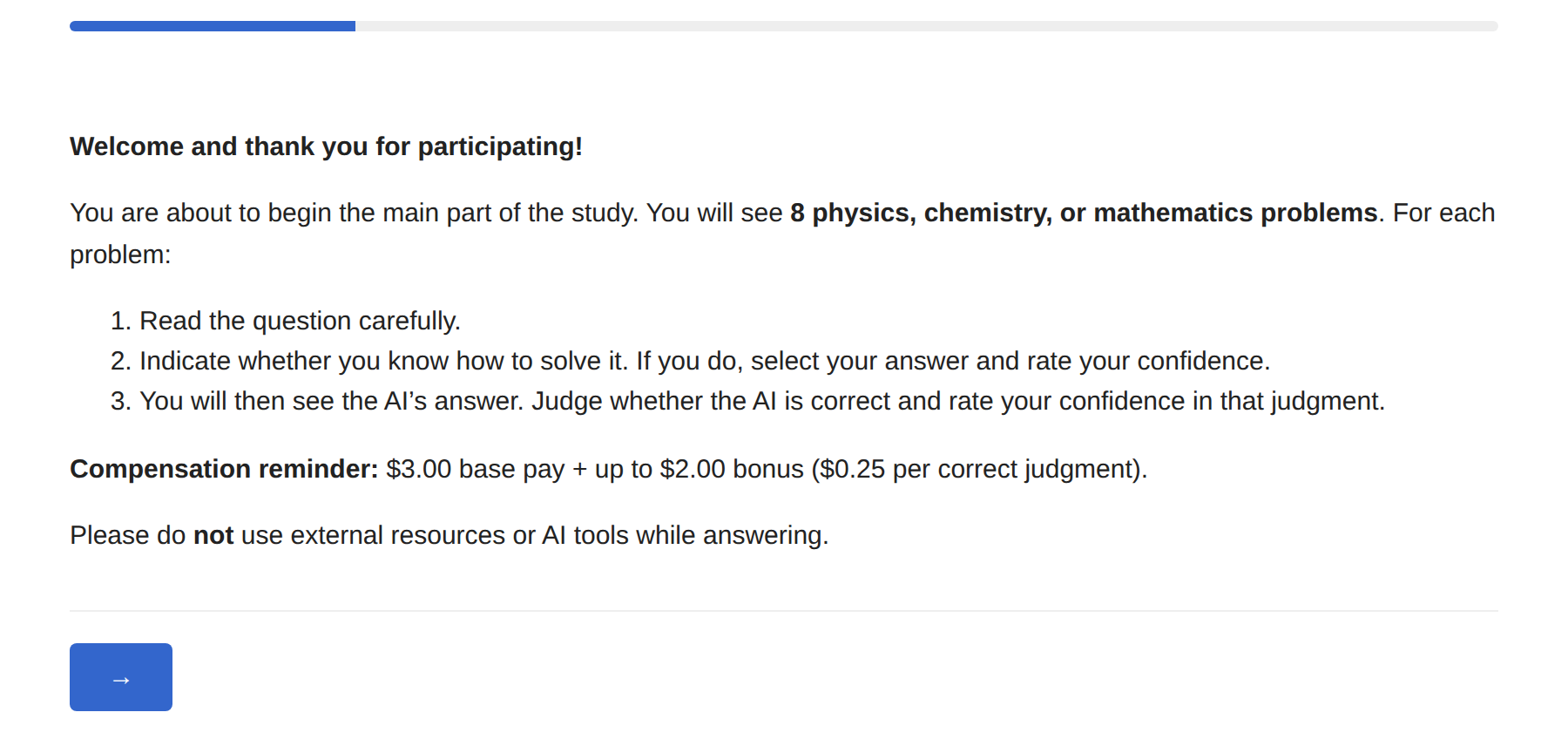}}
  \caption{\textbf{Step 4 -- Task instructions.} Describes the per-item flow
  the participant is about to begin (8 problems.}
  \label{fig:flow-instructions}
\end{figure}

\begin{figure}[ht!]
  \centering
  \fbox{\includegraphics[width=0.85\linewidth]{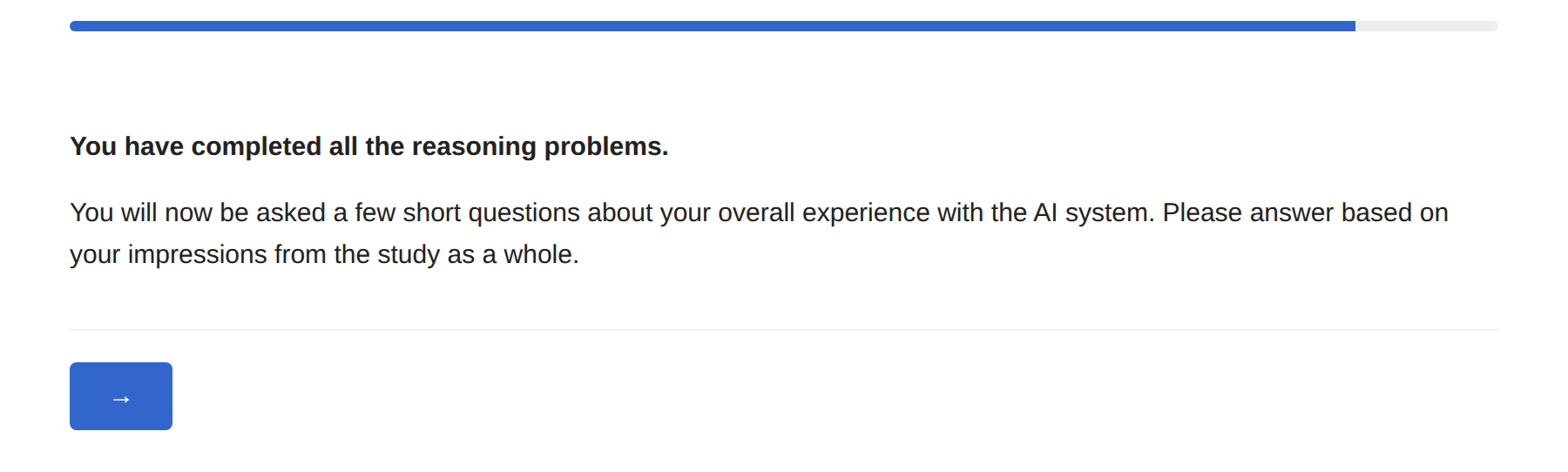}}
  \caption{\textbf{Step 8 -- End-of-task transition.} Confirms the participant
  has completed all 8 reasoning items and previews the post-study questionnaire.}
  \label{fig:flow-end}
\end{figure}

\begin{figure}[ht!]
  \centering
  \fbox{\includegraphics[width=0.85\linewidth]{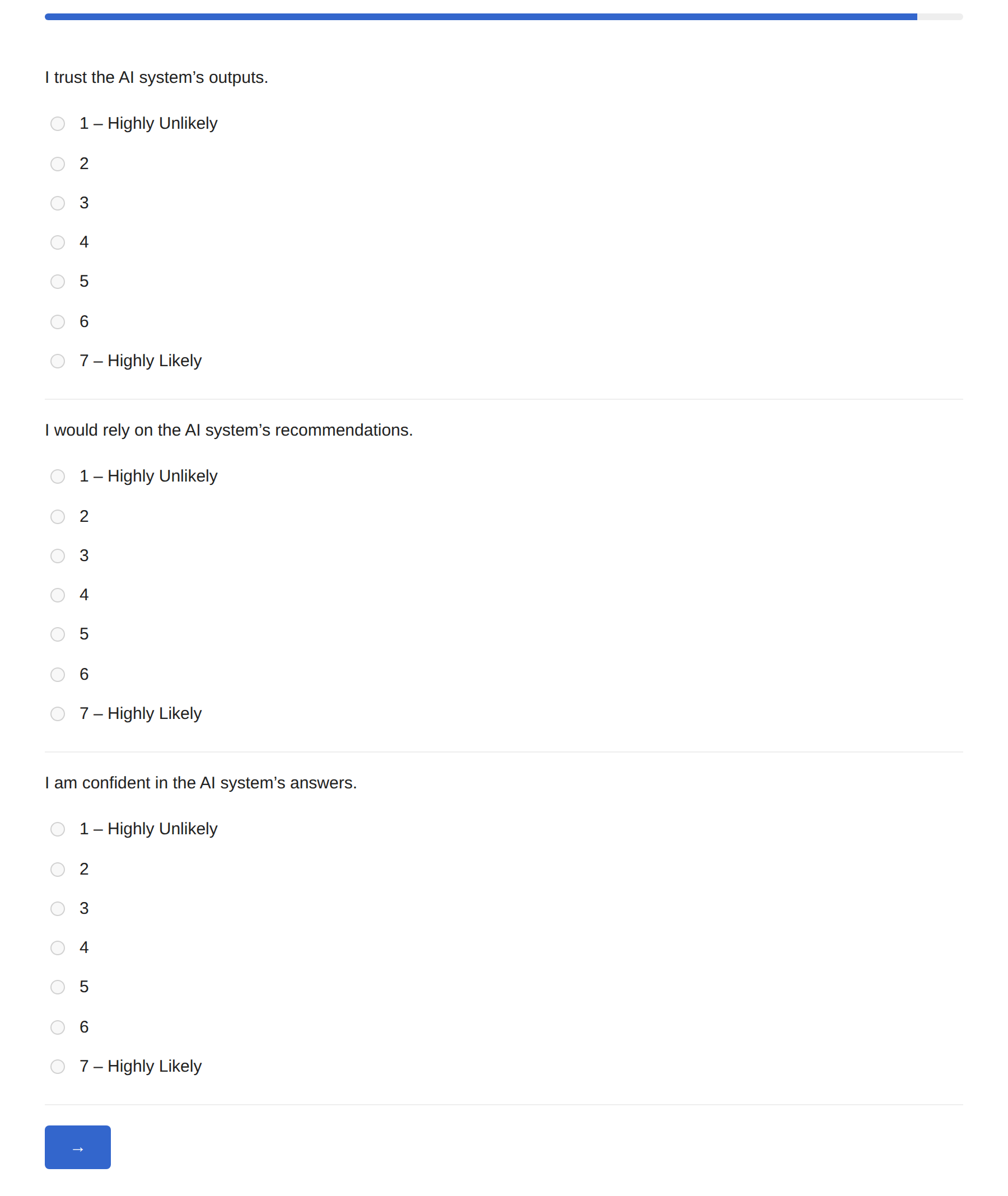}}
  \caption{\textbf{Step 9 -- Post-study trust questionnaire.} Three items
  rated on a 1--7 Likert scale (trust in outputs, willingness to rely on
  recommendations, confidence in answers).}
  \label{fig:flow-posttrust}
\end{figure}

\clearpage

\section{Statistical Analysis}
\label{app:stats}

\subsection{Confidence intervals}
\label{app:cis}

Table~\ref{tab:app-ci} reports Wilson 95\% CIs for every cell in the
main results tables. The intervals support the claims in
the main text: $E^{+/-}$ has the lowest 
trust ($40.9\%$, 95\% CI $[32.2,50.3]$) and misjudgment ($46.9\%$,
95\% CI $[37.2,56.8]$), and is tied for the highest accuracy
($60.4\%$, 95\% CI $[53.4,67.1]$).

\begin{table}[h]
\centering\small
\caption{Wilson 95\% confidence intervals on the metrics reported in
the main text. $k/n$ is shown for the relevant denominator
(Accuracy: all responses; False trust: responses judged correct;
Misjudgment: AI-incorrect responses; Correct judgment:
AI-correct responses).}
\label{tab:app-ci}
\begin{tabular}{l l c c}
\toprule
Metric & Condition & Rate (\%) & 95\% CI \\
\midrule
\multirow{5}{*}{Accuracy}
 & $E^{+/-}$            & 60.4 & [53.4, 67.1] \\
 & $E^{+}$              & 58.0 & [51.1, 64.6] \\
 & Reasoning summary    & 57.5 & [50.6, 64.1] \\
 & Reasoning trace      & 60.5 & [53.6, 67.0] \\
 & Solution only        & 53.1 & [46.1, 60.1] \\
\midrule
\multirow{5}{*}{False trust}
 & $E^{+/-}$            & 40.9 & [32.2, 50.3] \\
 & $E^{+}$              & 43.9 & [35.8, 52.5] \\
 & Reasoning summary    & 43.6 & [34.9, 52.6] \\
 & Reasoning trace      & 41.9 & [33.7, 50.5] \\
 & Solution only        & 47.3 & [38.2, 56.5] \\
\midrule
\multirow{5}{*}{Misjudgment}
 & $E^{+/-}$            & 46.9 & [37.2, 56.8] \\
 & $E^{+}$              & 58.0 & [48.2, 67.2] \\
 & Reasoning summary    & 51.0 & [41.3, 60.6] \\
 & Reasoning trace      & 54.0 & [44.3, 63.4] \\
 & Solution only        & 54.2 & [44.2, 63.8] \\
\midrule
\multirow{5}{*}{Correct judgment}
 & $E^{+/-}$            & 67.7 & [57.8, 76.2] \\
 & $E^{+}$              & 74.0 & [64.6, 81.6] \\
 & Reasoning summary    & 66.0 & [56.3, 74.5] \\
 & Reasoning trace      & 75.0 & [65.7, 82.5] \\
 & Solution only        & 60.4 & [50.4, 69.6] \\
\bottomrule
\end{tabular}
\end{table}

\subsection{Mixed-effects logistic regressions}
\label{app:glmm}
We fit a GLMM to estimate the effect of explanation condition while accounting for individual differences. We use$ E^{+/-}$ as the reference condition; results are reported in Table~\ref{tab:app-glmm}.

\begin{table}[h]
\centering\small
\caption{Mixed-effects logistic regression results. Odds ratios
(OR) and 95\% Wald CIs are relative to $E^{+/-}$
(reference). Each row tests a single condition against
$E^{+/-}$ for the specified outcome. Random intercepts for
participant and item. * denotes $p<.05$.}
\label{tab:app-glmm}
\begin{tabular}{l l c c c}
\toprule
Outcome & Comparator (vs.~$E^{+/-}$) & OR & 95\% CI & $p$ \\
\midrule
\multirow{4}{*}{Accuracy ($\uparrow$)}
 & $E^{+}$              & 0.89 & [0.66, 1.21] & .464 \\
 & Reasoning summary    & 0.88 & [0.65, 1.19] & .420 \\
 & Reasoning trace      & 1.01 & [0.75, 1.37] & .933 \\
 & Solution only        & 0.72 & [0.53, 0.97] & \textbf{.032*} \\
\midrule
\multirow{4}{*}{Misjudgment ($\downarrow$)}
 & $E^{+}$              & 1.73 & [1.11, 2.70] & \textbf{.017*} \\
 & Reasoning summary    & 1.17 & [0.76, 1.82] & .474 \\
 & Reasoning trace      & 1.40 & [0.90, 2.19] & .139 \\
 & Solution only        & 1.39 & [0.88, 2.17] & .156 \\
\midrule
\multirow{4}{*}{Correct judgment ($\uparrow$)}
 & $E^{+}$              & 1.48 & [0.91, 2.39] & .114 \\
 & Reasoning summary    & 0.92 & [0.58, 1.44] & .702 \\
 & Reasoning trace      & 1.57 & [0.96, 2.56] & .072 \\
 & Solution only        & 0.70 & [0.44, 1.09] & .112 \\
\midrule
\multirow{3}{*}{Helpfulness ($\uparrow$)}
 & $E^{+}$              & 2.80 & [1.80, 4.36] & \textbf{<.001*} \\
 & Reasoning summary    & 0.62 & [0.44, 0.88] & \textbf{.007*} \\
 & Reasoning trace      & 0.41 & [0.29, 0.58] & \textbf{<.001*} \\
\bottomrule
\end{tabular}
\end{table}

The mixed-effects analysis sharpens three claims from the main text:

\begin{enumerate}
\item \textbf{$E^{+/-}$ produces significantly higher accuracy than
the no-explanation baseline.} Solution only has $\sim$28\% lower odds
of a correct judgment than $E^{+/-}$ ($\mathrm{OR}=0.72$, 95\% CI
$[0.53,0.97]$, $p{=}.032$).

\item \textbf{$E^{+}$ induces significantly more misjudgment than
$E^{+/-}$.} Conditional on the AI being incorrect, the odds that a
user accepts the AI's answer are 1.7$\times$ higher under $E^{+}$
than under $E^{+/-}$ ($\mathrm{OR}=1.73$, 95\% CI $[1.11,2.70]$,
$p{=}.017$).

\item \textbf{$E^{+/-}$ is rated significantly more helpful than the
provenance-style explanations.} Compared with $E^{+/-}$, the odds of
a ``helpful'' rating are 0.62$\times$ for reasoning summary
($p{=}.007$) and 0.41$\times$ for reasoning trace ($p{<}.001$). The
only condition rated more helpful than $E^{+/-}$ is $E^{+}$
($\mathrm{OR}=2.80$), but $E^{+}$ achieves that helpfulness while
roughly doubling the odds of misjudgment.
\end{enumerate}

\subsection{Effect sizes for pairwise contrasts}
\label{app:effects}

Table~\ref{tab:app-effects} reports proportion differences with
Newcombe 95\% CIs and Cohen's $h$ with bootstrap 95\% CIs for the
two outcomes that drive our argument: misjudgment and accuracy.
Negative values for misjudgment and positive values for accuracy
favor $E^{+/-}$.

\begin{table}[h]
\centering\small
\caption{Pairwise contrasts vs. $E^{+/-}$. $\Delta p$ is the
proportion difference in percentage points (Newcombe 95\% CI);
$h$ is Cohen's $h$ with bootstrap percentile 95\% CI.}
\label{tab:app-effects}
\begin{tabular}{l l r r r r}
\toprule
Outcome & vs.~$E^{+/-}$ & $\Delta p$ (pp) & 95\% CI & $h$ & 95\% CI \\
\midrule
\multirow{4}{*}{Misjudgment}
 & $E^{+}$              & $-11.1$ & $[-24.5,+2.8]$ & $-0.22$ & $[-0.51,+0.06]$ \\
 & Reasoning summary    & $-4.1$  & $[-17.7,+9.7]$ & $-0.08$ & $[-0.37,+0.20]$ \\
 & Reasoning trace      & $-7.1$  & $[-20.6,+6.8]$ & $-0.14$ & $[-0.43,+0.14]$ \\
 & Solution only        & $-7.3$  & $[-20.9,+6.7]$ & $-0.15$ & $[-0.42,+0.13]$ \\
\midrule
\multirow{4}{*}{Accuracy}
 & $E^{+}$              & $+2.4$  & $[-7.3,+12.0]$ & $+0.05$ & $[-0.15,+0.25]$ \\
 & Reasoning summary    & $+2.9$  & $[-6.8,+12.5]$ & $+0.06$ & $[-0.13,+0.26]$ \\
 & Solution only        & $+7.3$  & $[-2.6,+17.0]$ & $+0.15$ & $[-0.05,+0.35]$ \\
\bottomrule
\end{tabular}
\end{table}

%% file: User_trust_exp.bbl
\begin{thebibliography}{51}
\providecommand{\natexlab}[1]{#1}
\providecommand{\url}[1]{\texttt{#1}}
\expandafter\ifx\csname urlstyle\endcsname\relax
  \providecommand{\doi}[1]{doi: #1}\else
  \providecommand{\doi}{doi: \begingroup \urlstyle{rm}\Url}\fi

\bibitem[Zhao et~al.(2026)Zhao, Zhou, Li, Tang, Wang, Hou, Min, Zhang, Zhang, Dong, Du, Yang, Chen, Chen, Jiang, Ren, Li, Tang, Liu, Liu, Nie, and Wen]{zhao2026}
Wayne~Xin Zhao, Kun Zhou, Junyi Li, Tianyi Tang, Xiaolei Wang, Yupeng Hou, Yingqian Min, Beichen Zhang, Junjie Zhang, Zican Dong, Yifan Du, Chen Yang, Yushuo Chen, Zhipeng Chen, Jinhao Jiang, Ruiyang Ren, Yifan Li, Xinyu Tang, Zikang Liu, Peiyu Liu, Jian-Yun Nie, and Ji-Rong Wen.
\newblock A survey of large language models, 2026.
\newblock URL \url{https://arxiv.org/abs/2303.18223}.

\bibitem[Campbell et~al.(2002)Campbell, Hoane, and Hsu]{deepblue}
Murray Campbell, A.~Joseph Hoane, and Feng-hsiung Hsu.
\newblock Deep blue.
\newblock \emph{Artif. Intell.}, 134\penalty0 (1–2):\penalty0 57–83, January 2002.
\newblock ISSN 0004-3702.
\newblock \doi{10.1016/S0004-3702(01)00129-1}.
\newblock URL \url{https://doi.org/10.1016/S0004-3702(01)00129-1}.

\bibitem[Silver et~al.(2017)Silver, Hubert, Schrittwieser, Antonoglou, Lai, Guez, Lanctot, Sifre, Kumaran, Graepel, Lillicrap, Simonyan, and Hassabis]{alphago}
David Silver, Thomas Hubert, Julian Schrittwieser, Ioannis Antonoglou, Matthew Lai, Arthur Guez, Marc Lanctot, Laurent Sifre, Dharshan Kumaran, Thore Graepel, Timothy Lillicrap, Karen Simonyan, and Demis Hassabis.
\newblock Mastering chess and shogi by self-play with a general reinforcement learning algorithm, 2017.
\newblock URL \url{https://arxiv.org/abs/1712.01815}.

\bibitem[Ribeiro et~al.(2016)Ribeiro, Singh, and Guestrin]{rib}
Marco~Tulio Ribeiro, Sameer Singh, and Carlos Guestrin.
\newblock "why should i trust you?": Explaining the predictions of any classifier, 2016.
\newblock URL \url{https://arxiv.org/abs/1602.04938}.

\bibitem[Doshi-Velez and Kim(2017)]{been}
Finale Doshi-Velez and Been Kim.
\newblock Towards a rigorous science of interpretable machine learning, 2017.
\newblock URL \url{https://arxiv.org/abs/1702.08608}.

\bibitem[Turpin et~al.(2023)Turpin, Michael, Perez, and Bowman]{turpin}
Miles Turpin, Julian Michael, Ethan Perez, and Samuel~R. Bowman.
\newblock Language models don't always say what they think: Unfaithful explanations in chain-of-thought prompting, 2023.
\newblock URL \url{https://arxiv.org/abs/2305.04388}.

\bibitem[Chen et~al.(2025)Chen, Benton, Radhakrishnan, Uesato, Denison, Schulman, Somani, Hase, Wagner, Roger, Mikulik, Bowman, Leike, Kaplan, and Perez]{anth}
Yanda Chen, Joe Benton, Ansh Radhakrishnan, Jonathan Uesato, Carson Denison, John Schulman, Arushi Somani, Peter Hase, Misha Wagner, Fabien Roger, Vlad Mikulik, Samuel~R. Bowman, Jan Leike, Jared Kaplan, and Ethan Perez.
\newblock Reasoning models don't always say what they think, 2025.
\newblock URL \url{https://arxiv.org/abs/2505.05410}.

\bibitem[Chua and Evans(2025)]{chua2025deepseek}
James Chua and Owain Evans.
\newblock Are deepseek r1 and other reasoning models more faithful?
\newblock In \emph{ICLR 2025 Workshop on Foundation Models in the Wild}, 2025.

\bibitem[Lanham et~al.(2023)Lanham, Chen, Radhakrishnan, Steiner, Denison, Hernandez, Li, Durmus, Hubinger, Kernion, et~al.]{lanham2023measuring}
Tamera Lanham, Anna Chen, Ansh Radhakrishnan, Benoit Steiner, Carson Denison, Danny Hernandez, Dustin Li, Esin Durmus, Evan Hubinger, Jackson Kernion, et~al.
\newblock Measuring faithfulness in chain-of-thought reasoning.
\newblock \emph{arXiv preprint arXiv:2307.13702}, 2023.

\bibitem[{DeepSeek-AI}(2025)]{deepseek}
{DeepSeek-AI}.
\newblock {DeepSeek-R1}: Incentivizing reasoning capability in {LLMs} via reinforcement learning.
\newblock \emph{Nature}, 645\penalty0 (8081):\penalty0 633--638, 2025.
\newblock \doi{10.1038/s41586-025-09422-z}.

\bibitem[Sreedharan et~al.(2024)Sreedharan, Kulkarni, and Kambhampati]{sree}
Sarath Sreedharan, Anagha Kulkarni, and Subbarao Kambhampati.
\newblock Explainable human-ai interaction: A planning perspective, 2024.
\newblock URL \url{https://arxiv.org/abs/2405.15804}.

\bibitem[Valmeekam et~al.(2025)Valmeekam, Stechly, Palod, Gundawar, and Kambhampati]{bs}
Karthik Valmeekam, Kaya Stechly, Vardhan Palod, Atharva Gundawar, and Subbarao Kambhampati.
\newblock Beyond semantics: The unreasonable effectiveness of reasonless intermediate tokens, 2025.
\newblock URL \url{https://arxiv.org/abs/2505.13775}.

\bibitem[Bhambri et~al.(2026)Bhambri, Biswas, and Kambhampati]{int}
Siddhant Bhambri, Upasana Biswas, and Subbarao Kambhampati.
\newblock Interpretable traces, unexpected outcomes: Investigating the disconnect in trace-based knowledge distillation, 2026.
\newblock URL \url{https://arxiv.org/abs/2505.13792}.

\bibitem[Kambhampati et~al.(2026)Kambhampati, Valmeekam, Bhambri, Palod, Saldyt, Stechly, Samineni, Kalwar, and Biswas]{kambhampati2026stop}
Subbarao Kambhampati, Karthik Valmeekam, Siddhant Bhambri, Vardhan Palod, Lucas Saldyt, Kaya Stechly, Soumya~Rani Samineni, Durgesh Kalwar, and Upasana Biswas.
\newblock Position: Stop anthropomorphizing intermediate tokens as reasoning/thinking traces!, 2026.

\bibitem[{OpenAI}(2024)]{openai}
{OpenAI}.
\newblock Learning to reason with {LLMs}.
\newblock \url{https://openai.com/index/learning-to-reason-with-llms/}, 2024.

\bibitem[{Gemini Team, Google}(2025)]{gemini}
{Gemini Team, Google}.
\newblock Gemini 2.5: Pushing the frontier with advanced reasoning, multimodality, long context, and next generation agentic capabilities, 2025.
\newblock URL \url{https://arxiv.org/abs/2507.06261}.

\bibitem[{Anthropic}(2025)]{anthropic}
{Anthropic}.
\newblock Claude's extended thinking.
\newblock \url{https://www.anthropic.com/news/visible-extended-thinking}, 2025.

\bibitem[{OpenAI}(2025)]{gptoss}
{OpenAI}.
\newblock gpt-oss-120b \& gpt-oss-20b model card, 2025.
\newblock URL \url{https://arxiv.org/abs/2508.10925}.

\bibitem[Fok and Weld(2024)]{fok}
Raymond Fok and Daniel~S. Weld.
\newblock In search of verifiability: Explanations rarely enable complementary performance in ai-advised decision making, 2024.
\newblock URL \url{https://arxiv.org/abs/2305.07722}.

\bibitem[Buçinca et~al.(2021)Buçinca, Malaya, and Gajos]{bucina}
Zana Buçinca, Maja~Barbara Malaya, and Krzysztof~Z. Gajos.
\newblock To trust or to think: Cognitive forcing functions can reduce overreliance on ai in ai-assisted decision-making.
\newblock \emph{Proceedings of the ACM on Human-Computer Interaction}, 5\penalty0 (CSCW1):\penalty0 1–21, April 2021.
\newblock ISSN 2573-0142.
\newblock \doi{10.1145/3449287}.
\newblock URL \url{http://dx.doi.org/10.1145/3449287}.

\bibitem[Arora et~al.(2023)Arora, Singh, and Mausam]{jee}
Daman Arora, Himanshu~Gaurav Singh, and Mausam.
\newblock Have llms advanced enough? a challenging problem solving benchmark for large language models, 2023.
\newblock URL \url{https://arxiv.org/abs/2305.15074}.

\bibitem[Si et~al.(2024)Si, Goyal, Wu, Zhao, Feng, Daum{\'e}~Iii, and Boyd-Graber]{si-etal-2024-large}
Chenglei Si, Navita Goyal, Tongshuang Wu, Chen Zhao, Shi Feng, Hal Daum{\'e}~Iii, and Jordan Boyd-Graber.
\newblock Large language models help humans verify truthfulness {--} except when they are convincingly wrong.
\newblock In Kevin Duh, Helena Gomez, and Steven Bethard, editors, \emph{Proceedings of the 2024 Conference of the North American Chapter of the Association for Computational Linguistics: Human Language Technologies (Volume 1: Long Papers)}, pages 1459--1474, Mexico City, Mexico, June 2024. Association for Computational Linguistics.
\newblock \doi{10.18653/v1/2024.naacl-long.81}.
\newblock URL \url{https://aclanthology.org/2024.naacl-long.81/}.

\bibitem[Steyvers et~al.(2025)Steyvers, Tejeda, Kumar, Belem, Karny, Hu, Mayer, and Smyth]{steyver}
Mark Steyvers, Heliodoro Tejeda, Aakriti Kumar, Catarina Belem, Sheer Karny, Xinyue Hu, Lukas~W. Mayer, and Padhraic Smyth.
\newblock What large language models know and what people think they know.
\newblock \emph{Nature Machine Intelligence}, 7\penalty0 (2):\penalty0 221–231, January 2025.
\newblock ISSN 2522-5839.
\newblock \doi{10.1038/s42256-024-00976-7}.
\newblock URL \url{http://dx.doi.org/10.1038/s42256-024-00976-7}.

\bibitem[Xiong et~al.(2024)Xiong, Hu, Lu, Li, Fu, He, and Hooi]{xiong}
Miao Xiong, Zhiyuan Hu, Xinyang Lu, Yifei Li, Jie Fu, Junxian He, and Bryan Hooi.
\newblock Can llms express their uncertainty? an empirical evaluation of confidence elicitation in llms, 2024.
\newblock URL \url{https://arxiv.org/abs/2306.13063}.

\bibitem[Ji et~al.(2023)Ji, Lee, Frieske, Yu, Su, Xu, Ishii, Bang, Madotto, and Fung]{ji}
Ziwei Ji, Nayeon Lee, Rita Frieske, Tiezheng Yu, Dan Su, Yan Xu, Etsuko Ishii, Ye~Jin Bang, Andrea Madotto, and Pascale Fung.
\newblock Survey of hallucination in natural language generation.
\newblock \emph{ACM Comput. Surv.}, 55\penalty0 (12), March 2023.
\newblock ISSN 0360-0300.
\newblock \doi{10.1145/3571730}.
\newblock URL \url{https://doi.org/10.1145/3571730}.

\bibitem[Manakul et~al.(2023)Manakul, Liusie, and Gales]{selfcheck}
Potsawee Manakul, Adian Liusie, and Mark J.~F. Gales.
\newblock Selfcheckgpt: Zero-resource black-box hallucination detection for generative large language models, 2023.
\newblock URL \url{https://arxiv.org/abs/2303.08896}.

\bibitem[Dahl et~al.(2024)Dahl, Magesh, Suzgun, and Ho]{Dahl_2024}
Matthew Dahl, Varun Magesh, Mirac Suzgun, and Daniel~E Ho.
\newblock Large legal fictions: Profiling legal hallucinations in large language models.
\newblock \emph{Journal of Legal Analysis}, 16\penalty0 (1):\penalty0 64–93, January 2024.
\newblock ISSN 1946-5319.
\newblock \doi{10.1093/jla/laae003}.
\newblock URL \url{http://dx.doi.org/10.1093/jla/laae003}.

\bibitem[Kim et~al.(2025)Kim, Vaughan, Liao, Lombrozo, and Russakovsky]{kim}
Sunnie S.~Y. Kim, Jennifer~Wortman Vaughan, Q.~Vera Liao, Tania Lombrozo, and Olga Russakovsky.
\newblock Fostering appropriate reliance on large language models: The role of explanations, sources, and inconsistencies.
\newblock In \emph{Proceedings of the 2025 CHI Conference on Human Factors in Computing Systems}, CHI ’25, page 1–19. ACM, April 2025.
\newblock \doi{10.1145/3706598.3714020}.
\newblock URL \url{http://dx.doi.org/10.1145/3706598.3714020}.

\bibitem[Sharma et~al.(2024)Sharma, Siu, Paleja, and Peña]{sharma}
Manasi Sharma, Ho~Chit Siu, Rohan Paleja, and Jaime~D. Peña.
\newblock Why would you suggest that? human trust in language model responses, 2024.
\newblock URL \url{https://arxiv.org/abs/2406.02018}.

\bibitem[Bo et~al.(2025)Bo, Wan, and Anderson]{bo}
Jessica~Y. Bo, Sophia Wan, and Ashton Anderson.
\newblock To rely or not to rely? evaluating interventions for appropriate reliance on large language models, 2025.
\newblock URL \url{https://arxiv.org/abs/2412.15584}.

\bibitem[Wang et~al.(2008)Wang, Jamieson, and Hollands]{inproceedings}
Lu~Wang, Greg Jamieson, and Justin Hollands.
\newblock Selecting methods for the analysis of reliance on automation.
\newblock volume~52, pages 287--291, 09 2008.
\newblock \doi{10.1177/154193120805200419}.

\bibitem[Schemmer et~al.(2022)Schemmer, Hemmer, Kühl, Benz, and Satzger]{schemmer}
Max Schemmer, Patrick Hemmer, Niklas Kühl, Carina Benz, and Gerhard Satzger.
\newblock Should i follow ai-based advice? measuring appropriate reliance in human-ai decision-making, 2022.
\newblock URL \url{https://arxiv.org/abs/2204.06916}.

\bibitem[Bansal et~al.(2021)Bansal, Wu, Zhou, Fok, Nushi, Kamar, Ribeiro, and Weld]{bansal}
Gagan Bansal, Tongshuang Wu, Joyce Zhou, Raymond Fok, Besmira Nushi, Ece Kamar, Marco~Tulio Ribeiro, and Daniel Weld.
\newblock Does the whole exceed its parts? the effect of ai explanations on complementary team performance.
\newblock In \emph{Proceedings of the 2021 CHI Conference on Human Factors in Computing Systems}, CHI '21, New York, NY, USA, 2021. Association for Computing Machinery.
\newblock ISBN 9781450380966.
\newblock \doi{10.1145/3411764.3445717}.
\newblock URL \url{https://doi.org/10.1145/3411764.3445717}.

\bibitem[Zhang et~al.(2020)Zhang, Liao, and Bellamy]{zhang}
Yunfeng Zhang, Q.~Vera Liao, and Rachel K.~E. Bellamy.
\newblock Effect of confidence and explanation on accuracy and trust calibration in ai-assisted decision making.
\newblock In \emph{Proceedings of the 2020 Conference on Fairness, Accountability, and Transparency}, FAT* '20, page 295–305, New York, NY, USA, 2020. Association for Computing Machinery.
\newblock ISBN 9781450369367.
\newblock \doi{10.1145/3351095.3372852}.
\newblock URL \url{https://doi.org/10.1145/3351095.3372852}.

\bibitem[Lai and Tan(2019)]{10.1145/3287560.3287590}
Vivian Lai and Chenhao Tan.
\newblock On human predictions with explanations and predictions of machine learning models: A case study on deception detection.
\newblock In \emph{Proceedings of the Conference on Fairness, Accountability, and Transparency}, FAT* '19, page 29–38, New York, NY, USA, 2019. Association for Computing Machinery.
\newblock ISBN 9781450361255.
\newblock \doi{10.1145/3287560.3287590}.
\newblock URL \url{https://doi.org/10.1145/3287560.3287590}.

\bibitem[Adadi and Berrada(2018)]{article}
Amina Adadi and Mohammed Berrada.
\newblock Peeking inside the black-box: A survey on explainable artificial intelligence (xai).
\newblock \emph{IEEE Access}, PP:\penalty0 1--1, 09 2018.
\newblock \doi{10.1109/ACCESS.2018.2870052}.

\bibitem[{OpenAI}(2023)]{gpt-4o-mini}
{OpenAI}.
\newblock {GPT-4} technical report, 2023.
\newblock URL \url{https://arxiv.org/abs/2303.08774}.

\bibitem[{DeepSeek-AI}(2024)]{v3}
{DeepSeek-AI}.
\newblock {DeepSeek-V3} technical report, 2024.
\newblock URL \url{https://arxiv.org/abs/2412.19437}.

\bibitem[Team(2025)]{qwq32b}
Qwen Team.
\newblock Qwq-32b: Embracing the power of reinforcement learning, March 2025.
\newblock URL \url{https://qwenlm.github.io/blog/qwq-32b/}.

\bibitem[{Qwen Team}(2024)]{qwen2.5_2024}
{Qwen Team}.
\newblock Qwen2.5 technical report.
\newblock \emph{arXiv preprint arXiv:2412.15115}, 2024.
\newblock URL \url{https://arxiv.org/abs/2412.15115}.

\bibitem[Jaech et~al.(2024)Jaech, Kalai, Lerer, Richardson, El-Kishky, Low, Helyar, Madry, Beutel, Carney, et~al.]{jaech2024openai}
Aaron Jaech, Adam Kalai, Adam Lerer, Adam Richardson, Ahmed El-Kishky, Aiden Low, Alec Helyar, Aleksander Madry, Alex Beutel, Alex Carney, et~al.
\newblock Openai o1 system card.
\newblock \emph{arXiv preprint arXiv:2412.16720}, 2024.

\bibitem[Singh et~al.(2025)Singh, Fry, Perelman, Tart, Ganesh, El-Kishky, McLaughlin, Low, Ostrow, Ananthram, et~al.]{singh2025openai}
Aaditya Singh, Adam Fry, Adam Perelman, Adam Tart, Adi Ganesh, Ahmed El-Kishky, Aidan McLaughlin, Aiden Low, AJ~Ostrow, Akhila Ananthram, et~al.
\newblock Openai gpt-5 system card.
\newblock \emph{arXiv preprint arXiv:2601.03267}, 2025.

\bibitem[Comanici et~al.(2025)Comanici, Bieber, Schaekermann, Pasupat, Sachdeva, Dhillon, Blistein, Ram, Zhang, Rosen, et~al.]{comanici2025gemini}
Gheorghe Comanici, Eric Bieber, Mike Schaekermann, Ice Pasupat, Noveen Sachdeva, Inderjit Dhillon, Marcel Blistein, Ori Ram, Dan Zhang, Evan Rosen, et~al.
\newblock Gemini 2.5: Pushing the frontier with advanced reasoning, multimodality, long context, and next generation agentic capabilities.
\newblock \emph{arXiv preprint arXiv:2507.06261}, 2025.

\bibitem[Nye et~al.(2021)Nye, Andreassen, Gur-Ari, Michalewski, Austin, Bieber, Dohan, Lewkowycz, Bosma, Luan, et~al.]{nye2021show}
Maxwell Nye, Anders~Johan Andreassen, Guy Gur-Ari, Henryk Michalewski, Jacob Austin, David Bieber, David Dohan, Aitor Lewkowycz, Maarten Bosma, David Luan, et~al.
\newblock Show your work: Scratchpads for intermediate computation with language models.
\newblock 2021.

\bibitem[Wei et~al.(2022)Wei, Wang, Schuurmans, Bosma, Xia, Chi, Le, Zhou, et~al.]{wei2022chain}
Jason Wei, Xuezhi Wang, Dale Schuurmans, Maarten Bosma, Fei Xia, Ed~Chi, Quoc~V Le, Denny Zhou, et~al.
\newblock Chain-of-thought prompting elicits reasoning in large language models.
\newblock \emph{Advances in neural information processing systems}, 35:\penalty0 24824--24837, 2022.

\bibitem[Muennighoff et~al.(2025)Muennighoff, Yang, Shi, Li, Fei-Fei, Hajishirzi, Zettlemoyer, Liang, Cand{\`e}s, and Hashimoto]{muennighoff2025s1}
Niklas Muennighoff, Zitong Yang, Weijia Shi, Xiang~Lisa Li, Li~Fei-Fei, Hannaneh Hajishirzi, Luke Zettlemoyer, Percy Liang, Emmanuel Cand{\`e}s, and Tatsunori Hashimoto.
\newblock s1: Simple test-time scaling.
\newblock \emph{arXiv preprint arXiv:2501.19393}, 2025.

\bibitem[Gandhi et~al.(2025)Gandhi, Chakravarthy, Singh, Lile, and Goodman]{gandhi2025cognitive}
Kanishk Gandhi, Ayush Chakravarthy, Anikait Singh, Nathan Lile, and Noah~D Goodman.
\newblock Cognitive behaviors that enable self-improving reasoners, or, four habits of highly effective stars.
\newblock \emph{arXiv preprint arXiv:2503.01307}, 2025.

\bibitem[Arcuschin et~al.(2025)Arcuschin, Janiak, Krzyzanowski, Rajamanoharan, Nanda, and Conmy]{arcuschin2025chain}
Iv{\'a}n Arcuschin, Jett Janiak, Robert Krzyzanowski, Senthooran Rajamanoharan, Neel Nanda, and Arthur Conmy.
\newblock Chain-of-thought reasoning in the wild is not always faithful.
\newblock \emph{arXiv preprint arXiv:2503.08679}, 2025.

\bibitem[Ouyang et~al.(2022)Ouyang, Wu, Jiang, Almeida, Wainwright, Mishkin, Zhang, Agarwal, Slama, Ray, Schulman, Hilton, Kelton, Miller, Simens, Askell, Welinder, Christiano, Leike, and Lowe]{rlhf}
Long Ouyang, Jeff Wu, Xu~Jiang, Diogo Almeida, Carroll~L. Wainwright, Pamela Mishkin, Chong Zhang, Sandhini Agarwal, Katarina Slama, Alex Ray, John Schulman, Jacob Hilton, Fraser Kelton, Luke Miller, Maddie Simens, Amanda Askell, Peter Welinder, Paul Christiano, Jan Leike, and Ryan Lowe.
\newblock Training language models to follow instructions with human feedback, 2022.
\newblock URL \url{https://arxiv.org/abs/2203.02155}.

\bibitem[Ibrahim et~al.(2025)Ibrahim, Hafner, and Rocher]{ibrahim}
Lujain Ibrahim, Franziska~Sofia Hafner, and Luc Rocher.
\newblock Training language models to be warm and empathetic makes them less reliable and more sycophantic, 2025.
\newblock URL \url{https://arxiv.org/abs/2507.21919}.

\bibitem[Chawla et~al.(2026)Chawla, Sokol, and Ganapini]{chawla2026llms}
Nitesh Chawla, Anna Sokol, and Marianna Ganapini.
\newblock Do llms have core beliefs?
\newblock 2026.

\end{thebibliography}
